\documentclass[a4paper,11pt]{article}
\pdfoutput=1

\usepackage{jcappub}
\usepackage[T1]{fontenc}
\usepackage{soul}
\usepackage[dvipsnames]{xcolor}
\usepackage{todonotes}
\usepackage[normalem]{ulem}

\usepackage{multicol}
\usepackage{multirow}

\newcommand{\nua}[1]{\ensuremath{\rlap{\kern-2.5pt\ensuremath{\overset{\scriptscriptstyle(-)}{\phantom{\nu}}}}{\ensuremath{{\nu}_{#1}}}}}
\newcommand{\Neff}{\ensuremath{N_\mathrm{eff}}}
\newcommand{\mnu}{\ensuremath{\Sigma m_\nu}}
\newcommand{\dmij}[1]{\ensuremath{\Delta m_{#1}^2}}
\newcommand{\Uaj}[1]{\ensuremath{|U_{#1}|^2}}
\newcommand{\ex}[1]{\ensuremath{\times10^{#1}}}
\newcommand{\nbb}{\ensuremath{0\nu\beta\beta}}

\title{\boldmath Bounds on light sterile neutrino mass and mixing from cosmology and laboratory searches}

\author[a,1]{Steffen Hagstotz,\note{Corresponding author.}}
\author[a]{Pablo F.\ de Salas,}
\author[b]{Stefano Gariazzo,}
\author[c,d]{Martina Gerbino,}
\author[d]{Massimiliano Lattanzi,}
\author[e]{Sunny Vagnozzi,}
\author[a,f,g]{Katherine Freese}
\author[b]{and Sergio Pastor}

\affiliation[a]{The Oskar Klein Centre for Cosmoparticle Physics, Department of Physics, Stockholm University, Roslagstullsbacken 21A, SE-106 91 Stockholm, Sweden}
\affiliation[b]{Instituto de F\'{i}sica Corpuscular (CSIC-Universitat de Val\`encia), Parc Cient\'{i}fic UV, c/ Cate\-dr\'{a}tico Jos\'{e} Beltr\'{a}n 2, E-46980 Paterna, Spain}
\affiliation[c]{High Energy Physics Division, Argonne National Laboratory, 9700 S. Cass Avenue, Lemont, IL 60439-4815, USA}
\affiliation[d]{Istituto Nazionale di Fisica Nucleare (INFN), Sezione di Ferrara, Via Giuseppe Saragat 1, I-44122 Ferrara, Italy}
\affiliation[e]{Kavli Institute for Cosmology (KICC) and Institute of Astronomy, University of Cambridge, Madingley Road, Cambridge CB3 0HA, United Kingdom}
\affiliation[f]{Theory Group, Department of Physics, The University of Texas at Austin, 2515 Speedway, C1600, Austin, TX 78712-0264, USA}
\affiliation[g]{Leinweber Center for Theoretical Physics, Department of Physics, University of Michigan, 450 Church Street, Ann Arbor, MI 48109, USA}

\emailAdd{steffen.hagstotz@fysik.su.se}
\emailAdd{pablo.fernandez@fysik.su.se}
\emailAdd{gariazzo@ific.uv.es}
\emailAdd{gerbinom@fe.infn.it}
\emailAdd{lattanzi@fe.infn.it}
\emailAdd{sunny.vagnozzi@ast.cam.ac.uk}
\emailAdd{ktfreese@umich.edu}
\emailAdd{pastor@ific.uv.es}

\abstract{
We provide a consistent framework to set limits on properties of light sterile neutrinos coupled to all three active neutrinos using a combination of the latest cosmological data and terrestrial measurements from oscillations, $\beta$-decay and neutrinoless double-$\beta$ decay ($\nbb$) experiments. We directly constrain the full $3+1$ active-sterile mixing matrix elements $\Uaj{\alpha4}$, with $\alpha \in ( e,\mu ,\tau )$, and the mass-squared splitting $\Delta m^2_{41} \equiv m_4^2-m_1^2$. We find that results for a $3+1$ case differ from previously studied $1+1$ scenarios where the sterile is only coupled to one of the neutrinos, which is largely explained by parameter space volume effects. Limits on the mass splitting and the mixing matrix elements are currently dominated by the cosmological data sets. The exact results are slightly prior dependent, but we reliably find all matrix elements to be constrained below $\Uaj{\alpha4}\lesssim 10^{-3}$. 

Short-baseline neutrino oscillation hints in favor of eV-scale sterile neutrinos are in serious tension with these bounds, irrespective of prior assumptions.
We also translate the bounds from the cosmological analysis into constraints on the parameters probed by laboratory searches, such as $m_\beta$ or $m_{\beta \beta}$, the effective mass parameters probed by $\beta$-decay and $\nbb$ searches, respectively. When allowing for mixing with a light sterile neutrino, cosmology leads to upper bounds of $m_\beta < 0.09$~eV and $m_{\beta \beta} < 0.07$~eV at 95\% C.L, more stringent than the limits from current laboratory experiments.
}

\begin{document}
\maketitle
\flushbottom

\section{Introduction}
\label{sec:intro}

The observation of oscillations between different neutrino flavors firmly establishes that at least two out of three neutrino mass eigenstates $m_i$ ($i=1,2,3$) are non-zero~\cite{Fukuda:1998mi,Ahmad:2002jz,Araki:2004mb,Adamson:2008zt,An:2012eh}. Since in the Standard Model of particle physics neutrinos only come with left-handed chirality, it is not possible to generate a mass term for them by the usual Higgs mechanism as for the other leptons. The discovery of neutrino masses then requires new physics beyond the Standard Model,
which may involve the existence of additional right-handed states (see e.g.,~\cite{Drewes:2013gca} for a review). These states are {\it sterile}, which means that they do not take part in weak, strong, or electromagnetic interactions, and their number and mass scale depend on the specific model.
For phenomenological purposes, in this paper we consider only one light sterile state with a mass at the eV scale~\cite{Giunti:2019aiy}.

One way to search for sterile neutrinos is in laboratory experiments. If produced in sufficient numbers in the early universe via oscillations with active states, sterile neutrinos can lead to observable effects in cosmology. As we will discuss in this paper, direct laboratory measurements and cosmological observations are complementary, and both avenues have led to tremendous progress in our understanding of the neutrino sector over the past few years.

Neutrino oscillation experiments have provided high-precision measurements of two mass-squared splittings \cite{deSalas:2017kay,Capozzi:2018ubv,Esteban:2018azc}:
\begin{align}
\dmij{21} &= m_2^2-m_1^2 = 7.55\ex{-5} \: \mathrm{eV}^2\,, \\
|\dmij{31}| &= |m_3^2-m_1^2| = 2.5\ex{-3} \: \mathrm{eV}^2 \,.
\end{align}
These mass splittings are responsible for the oscillations between the three neutrino flavor eigenstates associated with the charged 
leptons of the Standard Model.
Given the fact that the sign of $\dmij{31}$ is not determined, two mass orderings can be possibly realized in nature: the normal ordering, in which the lightest mass state is $m_1$ and $\dmij{31}>0$, or the inverted ordering, in which $m_3$ is the lightest state and $\dmij{31}<0$.
Current neutrino oscillation measurements show a mild statistical preference for the normal ordering~\cite{deSalas:2017kay,Capozzi:2018ubv,Esteban:2018azc,deSalas:2018bym}.

For many years, however, a number of experimental results that cannot be explained within the context of the three-neutrino oscillation paradigm have been reported.
Such anomalies are measured by various so-called short-baseline (SBL) experiments.
The first result was found by LSND~\cite{Aguilar:2001ty} and later partly confirmed by MiniBooNE \cite{Aguilar-Arevalo:2018gpe}, but anomalies were also detected by the Gallium experiments GALLEX and SAGE\footnote{GALLEX and SAGE were designed to find solar neutrinos, but the anomalies appear in the short-baseline calibration runs.}~\cite{Abdurashitov:2005tb,Giunti:2010zu,Kostensalo:2019vmv} and by a number of observations of reactor antineutrino fluxes at short distances~\cite{Mention:2011rk,Mueller:2011nm,Huber:2011wv}.
Although the situation is not completely clear, (see e.g.~\cite{Giunti:2019aiy,Boser:2019rta}), 
the anomalies point to an additional mass splitting, much larger than the other two:
\begin{equation}
\dmij{\rm SBL} \sim \mathcal O(\mathrm{eV}^2) \,.
\end{equation}
The existence of a new mass splitting implies the presence of a new neutrino mass eigenstate, which
corresponds to a fourth flavor eigenstate.
Since the hypothetical new neutrino has not been found in other weak interaction measurements~\cite{ALEPH:2005ab}, it cannot take part in Standard Model interactions and is therefore denoted ``sterile'' (see e.g.~\cite{Gariazzo:2015rra,Giunti:2019aiy}).
Due to the large gap $\mathcal O(\mathrm{eV}^2) \gg |\dmij{31}| , \, \dmij{21}$, oscillations with the sterile are not expected to affect standard atmospheric and solar neutrino measurements.
One typically assumes that the mixing between the three active neutrinos and the fourth mass eigenstate is small, so that the scenario is commonly referred to as the 3+1 model. For recent reviews on experimental searches for eV-scale sterile neutrinos, see also~\cite{Giunti:2019aiy,Diaz:2019fwt,Boser:2019rta}.

In case such a new neutrino exists, its presence can affect the evolution of the Universe as well. Even if the fourth neutrino does not interact with the other standard model particles,
it would be produced in the early Universe via oscillations with the active flavors. Excluding direct detection of relic neutrinos~\cite{Betti:2019ouf}, the cosmological presence of the sterile state can only be deduced indirectly, for instance from its contribution to the energy density of the Universe~\cite{Abazajian:2017tcc}.
In particular, considering masses around the $\sim \mathrm{eV}$ scale as motivated by the SBL anomalies, the sterile neutrino is produced as a relativistic particle. Therefore, it contributes at early times to the number of relativistic degrees of freedom \Neff, which quantifies the amount of energy density from relativistic species beyond photons in the early Universe.
Moreover, the free-streaming length associated to light sterile neutrinos is potentially large, so that they would behave as hot dark matter and leave a distinct imprint on structure formation, similar to the one of active neutrinos~\cite{Hannestad:2010kz,Wong:2011ip,Lesgourgues:2014zoa,Lattanzi:2017ubx}. Cosmological observations are sensitive to the hot dark matter energy density $\Omega_\mathrm{hot} =(\Omega_\nu+\Omega_s) $, neglecting the contribution of other non-standard particles.
Recent measurements of the temperature and polarisation anisotropies in the cosmic microwave background (CMB) and of large-scale structures (LSS) are able to constrain both $\Neff$ and $\Omega_\mathrm{hot} h^2$ to high precision, see e.g.~\cite{Palanque-Delabrouille:2015pga,Cuesta:2015iho,Huang:2015wrx,Giusarma:2016phn,Vagnozzi:2017ovm,Wang:2017htc,Chen:2017ayg,Upadhye:2017hdl,Nunes:2017xon,Giusarma:2018jei,Choudhury:2018byy,Aghanim:2018eyx,Gariazzo:2018meg,RoyChoudhury:2019hls,Gariazzo:2019xhx}. Predictions of big-bang nucleosynthesis (BBN) combined with measurements of primordial element abundances also provide independent, albeit less stringent, constraints on $\Neff$ in agreement with CMB+LSS findings~\cite{Consiglio:2017pot}.


There is another reason to consider additional relativistic particles such as light sterile neutrinos in cosmology, independent of the anomalous results from neutrino experiments. Contributions to the energy density in the early Universe reduce the size of the sound horizon and result in a higher value of the Hubble constant $H_0$ inferred from baryon acoustic oscillations (see e.g. discussions in~\cite{Bernal:2016gxb,Lemos:2018smw,Poulin:2018cxd,Vagnozzi:2019ezj,Verde:2019ivm}). Currently measurements based on observing the sound horizon in the CMB or in the galaxy distribution at late times are in tension with local distance ladder measurements from supernovae Ia \cite{Riess:2016jrr,Riess:2018byc,Riess:2019cxk} and lensed quasar time delay measurements~\cite{Birrer:2018vtm,Chen:2019ejq,Wong:2019kwg}, see e.g.~\cite{Addison:2017fdm,Aylor:2018drw,Knox:2019rjx} for further discussions. However, a sterile neutrino alone cannot account for the full discrepancy, as discussed in~\cite{Aghanim:2018eyx}.

In this work, we study the phenomenology of a light sterile neutrino from various points of view. While the main goal is to compute, for the first time, cosmological bounds on all the mixing angles and the mass splitting between the light sterile state and active neutrinos, we also compare the cosmological results with other probes, such as constraints from direct mass measurements and neutrinoless double-$\beta$ decay, and limits from neutrino oscillation experiments. We will focus on light sterile neutrinos (i.e., $\dmij{41}<100\,\mathrm{eV^2}$), since we are motivated by anomalies in oscillation experiments that could be explained by the existence of light sterile states.

The rest of this paper is structured as follows. First, we give a brief overview on the theoretical aspects and the status of sterile neutrino probes in section~\ref{sec:ster_osc}.
The phenomenology of such neutrinos in the early and late Universe is discussed in section~\ref{sec:sterile_cosmology}. In section~\ref{sec:data} we discuss the datasets we consider and the method used to obtain the bounds on the sterile neutrino mixing parameters, for which we present upper limits in section~\ref{sec:results}. We summarize and conclude in section~\ref{sec:conclusion}.

\section{Experimental searches for light sterile neutrinos}
\label{sec:ster_osc}

In this section we briefly summarize the theory of active-sterile neutrino oscillations and review the status of light sterile neutrino searches at various experiments. For comprehensive reviews focused on experimental searches for eV-scale sterile neutrinos, see~\cite{Giunti:2019aiy,Diaz:2019fwt,Boser:2019rta}.

\subsection{Active-sterile neutrino mixing}
The mixing between active and sterile neutrinos can be parameterized by means of the unitary $4\times4$ lepton mixing matrix $U$. There are several ways to write the mixing matrix, so we will briefly explain our choice of parameterization. In the case of mixing between four neutrino states, $U$ is fully characterized by six mixing angles and three CP-violating Dirac phases, but for the purposes of this work we set the phases to zero and therefore assume that oscillations among neutrinos are identical to those among antineutrinos.
Three of the angles characterize the oscillations between the three active neutrinos: $\theta_{13}$, $\theta_{23}$, and $\theta_{23}$.
The remaining three mixing angles, namely, $\theta_{14}$, $\theta_{24}$, and $\theta_{34}$, describe the mixing with the sterile state.
We choose the following parameterization of the mixing matrix~\cite{Gariazzo:2015rra}:
\begin{equation}\label{eq:mixpar}
U = R^{34} R^{24} R^{14} R^{23} R^{13} R^{12} \,,
\end{equation}
where each $R^{ij}$ is a real matrix describing a rotation by an angle $\theta_{ij}$. 

Bounds on the mixing matrix can be provided in a way that is independent of the parametrization if instead of quoting limits on the individual mixing angles $\theta_{ij}$ we consider the matrix elements directly. Concerning active-sterile neutrino mixing, the most important elements are those of the fourth column:
\begin{align}
\Uaj{e4} &= \sin^2 \theta_{14} \,, \\
\Uaj{\mu 4} &= \cos^2 \theta_{14} \sin^2 \theta_{24} \,, \\
\Uaj{\tau 4} &= \cos^2 \theta_{14} \cos^2 \theta_{24} \sin^2 \theta_{34} \,, \\
\Uaj{s 4} &= \cos^2 \theta_{14} \cos^2 \theta_{24} \cos^2 \theta_{34} \,.
\end{align}
We expect the mixing angles $\theta_{i4}$ to be small, in order not to substantially alter the phenomenology of three-neutrino mixing beyond what is allowed by current limits. Therefore, the matrix elements \Uaj{e4}, \Uaj{\mu 4} and \Uaj{\tau 4} are expected to be small, while \Uaj{s 4} should in principle be of order unity. For this reason, it is possible to refer to the fourth neutrino mass eigenstate as the ``sterile'' neutrino, even though strictly speaking a sterile neutrino is a flavor eigenstate and has no definite mass. In the rest of the paper, we will often refer to the mass $m_s\simeq m_4$ as the mass of the sterile neutrino.

In addition to the mixing angles, we need to specify the three mass-squared splittings \dmij{21}, \dmij{31}, and \dmij{41}. For the active neutrinos, we always assume normal ordering\footnote{While not yet conclusive, recent cosmological and oscillations probes slightly favor normal ordering over inverted ordering, with varying degree of preference~\cite{Hannestad:2016fog,Xu:2016ddc,Gerbino:2016ehw,Vagnozzi:2017ovm,Schwetz:2017fey,deSalas:2017kay,Gariazzo:2018pei,deSalas:2018bym,RoyChoudhury:2019hls}.} ($\dmij{31}>0$) and we fix all the standard mixing parameters to the best-fit values obtained in \cite{deSalas:2017kay}.
When considering the full oscillation pattern, the complete expressions for the oscillation probabilities are rather involved and depend on all the mass splittings and mixing angles, see e.g.~\cite{Giunti:2007ry}. When considering SBL oscillations, however, the terms due to the much smaller solar and atmospheric mass-squared differences are suppressed because they correspond to slower oscillations, and only the effect of \dmij{41} is relevant. As a consequence, the oscillation probabilities can be well approximated by a two-neutrino mixing formula with appropriate mixing matrix elements. This is the appearance probability to populate a flavor state:
\begin{equation}
P^{\mbox{SBL}}_{\nu_{\alpha}\rightarrow\nu_{\beta}} \simeq \sin^2 \left ( 2\vartheta_{\alpha\beta} \right ) \sin^2\left(\frac{\dmij{41}L}{4E}\right)\,,
\label{eq:P_sbl_app}
\,
\qquad(\alpha\neq\beta)
\end{equation}
whereas the flux of the initial flavor is modulated by the disappearance probability:
\begin{equation}
P^{\mbox{SBL}}_{\nu_{\alpha}\rightarrow\nu_{\alpha}} \simeq 1 - \sin^2 \left ( 2\vartheta_{\alpha\alpha} \right )\sin^2\left(\frac{\dmij{41}L}{4E}\right)
\label{eq:P_sbl_dis}
\,,
\end{equation}
where $L$ is the distance traveled by the neutrino, $E$ is its energy, and we use Greek letters to refer to the four flavor states $\alpha, \beta \in [e, \mu, \tau, s]$. The effective angles $\vartheta_{\alpha\alpha}$ and $\vartheta_{\alpha\beta}$ depend on the elements of the fourth column of the mixing matrix:
\begin{eqnarray} \sin^2 \left ( 2\vartheta_{\alpha\beta} \right ) &=& 4\,\Uaj{\alpha4}\,\Uaj{\beta4} \,,
\qquad(\alpha\neq\beta)
\label{eq:theta_eff_app}
\\
\sin^2 \left ( 2\vartheta_{\alpha\alpha} \right ) &=& 4\,\Uaj{\alpha4}\, (1- \Uaj{\alpha4})
\label{eq:theta_eff_dis}
\,.
\end{eqnarray}
The last expression is implied by the unitarity of the mixing matrix.  From the probabilities in Eqs.~(\ref{eq:P_sbl_app}) and~(\ref{eq:P_sbl_dis}) it is clear why the oscillatory behavior manifests when $4E\sim\Delta m^2 L$. If $4E\gg\Delta m^2 L$, the oscillatory term vanishes; if $4E\ll\Delta m^2 L$, the oscillation frequency becomes so high that oscillations cannot be resolved anymore and they are averaged out. The eV scale we focus on arises since various experiments find anomalies pointing towards $(L/E)^{-1} \sim \mathcal O (\mathrm{eV}^{2})$. We emphasize that the absolute mass scale of the neutrino states does not enter in the equations, so oscillation experiments depend on the differences of the mass-squared values alone. In other words, oscillations are independent of the lightest neutrino mass $m_1$. Both appearance and disappearance channels can be used to measure the effective mixing angles and to constrain the mixing matrix elements.

\subsection{Current status of oscillation data and global fits}
\label{subsec:global_fits}

There is not a single SBL anomaly, but several different experiments with short baselines find anomalous neutrino fluxes of varying degree. We will briefly summarize the current situation in this section.

From the historical point of view, the first anomaly was found by
LSND \cite{Aguilar:2001ty},
which reported the unexpected appearance of electron antineutrinos
in a beam of muon antineutrinos produced from $\pi^+$ decays.
Years later, also
MiniBooNE \cite{Aguilar-Arevalo:2018gpe} confirmed the excess of $\bar \nu_e$ events, with a similar experimental setup, in partial agreement with LSND even though the MiniBooNE excess is too large to be explained by a sterile alone.

Another anomaly with comparable $L/E$ was found by $\nu_e$ disappearance measurements by the Gallium neutrino detectors GALLEX and SAGE \cite{Abdurashitov:2005tb,Abdurashitov:2009tn,Giunti:2010zu,Kaether:2010ag}. Both experiments measured the electron neutrino flux in proximity to a radioactive source, and found a lack of events at significance of $\sim 3 \sigma$. A similar effect is observed in measurements of $\bar \nu_e$ disappearance in close proximity to nuclear reactors \cite{Mention:2011rk}. The lack of electron antineutrinos also reaches a significance of $\sim 3 \sigma$ and was noticed after new calculations found a higher expected initial flux \cite{Mueller:2011nm,Huber:2011wv}.

The LSND and MiniBooNE $\bar \nu_e$ appearance data require a non-zero value of $\vartheta_{e \mu}$, while in order to explain the reactor and Gallium disappearance measurements one needs $\vartheta_{ee} > 0$. Taken together, this also implies a non-zero mixing $\vartheta_{\mu \mu}$ and $\Uaj{\mu 4}$.
However, this matrix element can be measured independently by muon (anti) neutrino disappearance experiments, and no corresponding anomaly for the relevant $L/E$ values is found by either measurements of the atmospheric muon neutrino flux by IceCube \cite{Aartsen:2017bap,TheIceCube:2016oqi} or by the MINOS+ \cite{Adamson:2017uda} collaboration using an accelerator beam.
Therefore, a combination of $\bar \nu_e$ appearance data by LSND and MiniBooNE and the disappearance results from electron and muon (anti) neutrinos in a global fit is currently problematic \cite{Dentler:2018sju,sterile19}.

\subsection{Laboratory searches for the absolute neutrino mass scale}
\label{sec:lab_numass}

In addition to neutrino flavor oscillation experiments, laboratory searches for massive neutrinos also include kinematic measurements of $\beta$-decay and searches for neutrinoless double-$\beta$ ($\nbb$) decay events. In this section, we briefly summarize the consequences that a light sterile neutrino mixing with the active flavors can have for these searches.

One way to probe the absolute neutrino mass scale directly is to observe the cutoff of the electron energy spectrum emitted from $\beta$-decay. It occurs at the effective electron neutrino mass $m_\beta$, given by the incoherent sum of squared masses and mixing parameters:
\begin{equation}
\label{eq:m_b}
    m_\beta^2 = \sum_{j=1}^{4} |U_{ej}|^2 m_{j}^2
\end{equation}
where $j=4$ is the contribution from the additional sterile neutrino. So far, $\beta$-decay measurements have only been able to set upper limits on the mass scale, with the latest bound $m_\beta<1.1\,\mathrm{eV}$ at 90\% confidence level (C.L.) recently published by the 
KATRIN collaboration~\cite{Aker:2019uuj}.

Neutrinoless double-$\beta$ decay ($0\nu\beta\beta$) searches constrain the half-life $T_{1/2}$ of the isotope involved in the decay (see e.g.~\cite{DellOro:2016tmg} for a recent review). Assuming that the mechanism responsible for lepton number violation manifested in $0\nu\beta\beta$ events is the mass mechanism, constraints on the half life can be translated into constraints on the effective Majorana mass $m_{\beta\beta}$:

\begin{equation}
\label{eq:halflife}
    T_{1/2} = \frac{m_e^2}{G_{0 \nu} | M_{0 \nu}|^2 m_{\beta \beta}^2},
\end{equation}
where $m_e$ is the electron mass, $G_{0 \nu}$ is the phase space factor and $M_{0\nu}$ is the nuclear transition matrix element for the decay. The effective mass parameter $m_{\beta \beta}$ can be expressed as a coherent sum of mass eigenstates and mixing matrix parameters:

\begin{equation}
\label{eq:m_bb}
    m_{\beta \beta} = \left| \sum_{j=1}^4 |U_{ej}|^2 \mathrm{e}^{i \alpha_j} m_{j} \right| \,,
\end{equation}
where $\alpha_j$ are Majorana phases and $j=4$ again contains the contribution from the sterile neutrino. So far, no such event has been detected and only upper limits on the lifetime $T_{1/2}$ of various isotopes are available, as described in section~\ref{subsec:decaydata}. These bounds are usually converted into a range of upper limits on the Majorana mass $m_{\beta \beta}$ depending on theoretical uncertainty in the calculation of the nuclear matrix elements.

\section{Light sterile neutrinos in cosmology}
\label{sec:sterile_cosmology}

As explained in section~\ref{sec:intro}, sterile neutrinos do not take part in weak, strong or electromagnetic interactions. As a consequence, they will not be produced by Standard Model scattering or annihilations in the very early Universe. In this work, we consider a production mechanism via non-resonant oscillations with active states, the so-called Dodelson-Widrow production mechanism~\cite{Dodelson:1993je,Colombi:1995ze}. 
In the very early universe, densities are high and weak interactions frequent. This generates an effective matter potential that suppresses neutrino oscillations. Therefore the sterile state is only populated once densities are low enough for oscillations with the active neutrino eigenstates to occur (see e.g.~\cite{2013neco.book.....L} for a detailed discussion).

The mass splitting sets the timescale for oscillations with the fourth neutrino and determines the time when flavor oscillations can start to arise. Here we focus on eV-scale sterile neutrinos and consequently consider mass splittings $10^{-2}\leq\dmij{41}/\text{eV}^2\leq10^2$. This range corresponds to plasma temperatures between $\mathcal{O}(100)$ and $\mathcal{O}(1)$~MeV when the sterile starts to be populated.

The thermalization process is described by a set of differential equations
that encode the evolution of the momentum distribution functions $f_\alpha(p)$ of the various neutrino flavors $\alpha \in [e,\mu,\tau,s]$,
and an additional one for the evolution of the photon temperature, as described in detail in~\cite{Gariazzo:2019gyi}.
As far as cosmological effects of neutrinos are concerned, what matters is the momentum distribution function of all the neutrino states
after their decoupling from the thermal plasma and at the end of electron-positron annihilations into photons,
which happens at temperatures around $0.1$~MeV.

The final momentum distribution function for the active neutrinos is very close, but not exactly equal,
to a Fermi-Dirac shape with the neutrino temperature $T_\nu$ 
(see e.g.\ \cite{Dolgov:2002wy,Mangano:2005cc,deSalas:2016ztq,Gariazzo:2019gyi,Bennett:2019ewm}). 
The deviation from thermal equilibrium is mostly due to the fact that neutrino decoupling does not occur instantaneously, so there are small distortions at high momenta that come from the energy transferred
during electron-positron annihilations to the few neutrinos still coupled to the plasma.
For the sterile neutrino, the momentum distribution function $f_s$ depends on the degree of thermalization that it reaches.
Initially the sterile is absent, and if the mass splitting and the mixing angles are not large enough, oscillations either start too late or are not efficient enough to bring the fourth neutrino into equilibrium with the active flavors.
The degree of thermalization of the sterile can be expressed in terms of the effective number
of relativistic species ($\Neff$), which can be constrained by big-bang nucleosynthesis (BBN)~\cite{Consiglio:2017pot} 
and more tightly by CMB observations~\cite{Aghanim:2018oex}. After electron-positron annihilations, this parameter can be expressed as
\begin{equation}
\Neff
=
\frac{8}{7}
\left(\frac{11}{4}\right)^{4/3}
\frac{(\rho_\nu+\rho_s)}{\rho_\gamma}\:,
\end{equation}
with the energy density in photons $\rho_\gamma$ and active plus sterile neutrinos $(\rho_\nu+\rho_s)$. They can be computed from the integrated distribution functions, and for negligible contributions from the sterile we recover the standard value $\Neff^{3\nu}=3.043-3.045$~\cite{Mangano:2005cc,deSalas:2016ztq,Gariazzo:2019gyi,Bennett:2019ewm,Escudero:2020dfa}.
On the other hand, if the mixing parameters are sufficiently large,
there is time for neutrino oscillations to fully bring the fourth neutrino
to equilibrium with the active flavors.
In such case, the final distribution function of sterile neutrinos $f_s$ will also be very close to a Fermi-Dirac spectrum,
with the same temperature as the active neutrinos,
and $\Neff\simeq4.05$.
Intermediate cases correspond to an incomplete thermalization,
the sterile neutrino contributes with $\Delta\Neff=\Neff-\Neff^{3\nu}<1.01$
and its distribution function is significantly non-thermal.
In such cases, however, it turns out that since the sterile is populated through oscillations from the thermal active states, the distribution function $f_s$ is still proportional to a Fermi-Dirac shape~\cite{Gariazzo:2019gyi} with temperature $T_\nu$. Then, we generally have \cite{Dodelson:1993je,Colombi:1995ze}
\begin{equation}
\label{eq:f_s}
f_s(p) = \frac{\Delta\Neff}{\exp (p/T_\nu) + 1} \: ,
\end{equation}
where $p$ is the neutrino momentum
and $T_\nu$ is the temperature of active neutrinos.
Note that for all cases considered here,
the sterile neutrino distribution function reaches its asymptotic values
well before primordial BBN at $T\sim 0.1 \: \mathrm{MeV}$ \cite{Gariazzo:2019gyi}.
This means that, under our assumptions,
the value of $\Neff$ does not change between the time of BBN and CMB decoupling, which would have consequences for the abundance of light elements~\cite{Consiglio:2017pot} important for the calculation of cosmological perturbations
(see also \cite{Schoneberg:2019wmt}).

The problem of active-sterile oscillations in the early Universe has been studied in many previous papers, some of them published more than 30 years 
ago (early references include e.g.\ 
\cite{Barbieri:1989ti,Kainulainen:1990ds,Barbieri:1990vx,Enqvist:1990ad,Enqvist:1991qj}, 
see the review \cite{Dolgov:2002wy} for an extensive list). Solving the Boltzmann kinetic equations for different neutrino energies is a complex issue, due to the simultaneous presence of neutrino interactions via weak processes and flavor oscillations in an expanding medium. Thus, past analyses 
\cite{Abazajian:2001nj,Dolgov:2003sg,Cirelli:2004cz,Hannestad:2012ky,Hannestad:2013wwj,Hannestad:2015tea,Mirizzi:2012we,Mirizzi:2013gnd,Saviano:2013ktj,Bridle:2016isd,Knee:2018rvj,Berryman:2019nvr,Adams:2020nue}
have considered various approaches that approximated the multi-momentum calculations 
to the evolution of an average momentum and/or reduced the number of active neutrino states.
In particular, the \texttt{LASAGNA} code \cite{Hannestad:2013wwj} solves the quantum kinetic equations in the 
1+1 case (assuming one active coupled via oscillations to one sterile neutrino state) with full collision integrals.
This code has been used in previous works 
\cite{Hannestad:2012ky,Bridle:2016isd,Knee:2018rvj,Berryman:2019nvr}
to map the active-sterile mixing parameters onto two other quantities relevant for cosmology (\Neff\ and the effective sterile neutrino mass).

When considering a simplified model with only one active neutrino $\nu_a$ and one sterile $\nu_s$, it is possible to relate the degree of thermalization $\Delta \Neff$ directly to the mixing parameters. For a mass splitting $\delta m^2_{as}$ and mixing angle $\vartheta_{as}$, this results in \cite{Dolgov:2003sg}
\begin{equation}\label{eq:deltaneff_mixing_approx}
\frac{\delta m^2_{as}}{\text{eV}^2}
\sin^4\left(2\vartheta_{as}\right)
\simeq
10^{-5}
\ln^2\left(1-\Delta\Neff\right) \: ,
\end{equation}
where the numerical coefficient is slightly different for electron, muon, or tau flavor neutrinos.
If one applies this relation to the 3+1 case,
with $\delta m^2_{as}\rightarrow\dmij{41}$
and
$\sin^2\left(2\vartheta_{as}\right)\simeq4\Uaj{\alpha4}\Uaj{s4}$,
one gets that in order to have a fully thermalized sterile neutrino
with a mass splitting around 1~eV$^2$,
a mixing matrix element $\Uaj{\alpha4}\simeq10^{-3}$ is required.
From this relation we can also see that a larger mixing matrix element generally increases \Neff\ towards 4. For larger mass splittings, a smaller mixing is sufficient to generate the same level of thermalization since the oscillations are faster, see Eq.~(\ref{eq:P_sbl_app})-(\ref{eq:P_sbl_dis}).
While Eq.~(\ref{eq:deltaneff_mixing_approx}) is a rough estimate, we will see that it is a quite good approximation of the full calculation,
as long as one mixing angle is varied at each time.

Including also the mixing among the active neutrino states has also been considered before, from the early simplified analyses \cite{Dolgov:2003sg,Cirelli:2004cz} to more recent multi-angle studies \cite{Mirizzi:2012we,Mirizzi:2013gnd,Adams:2020nue} performed within the averaged-momentum approximation. Although some authors have studied the multi-angle (2+1 scenario) and multi-momentum problem \cite{Saviano:2013ktj}, only very recently the evolution in the early Universe of the momentum-dependent kinetic equations for the full $4\times4$ density matrix of neutrinos was calculated with
a dedicated numerical code, \texttt{FortEPiaNO}~\footnote{\url{https://bitbucket.org/ahep_cosmo/fortepiano_public}}, as
described in~\cite{Gariazzo:2019gyi}.

At late times, the sterile neutrino becomes non-relativistic and it contributes to the matter energy density.
At such point in the evolution, 
its contribution to the energy density is~\cite{2013neco.book.....L,Colombi:1995ze}
\begin{equation}
\label{eq:Omega_s}
\Omega_s h^2 = \frac{m_s \Delta \Neff}{93.14~\mathrm{eV}}\equiv \frac{m_{s,\,\mathrm{eff}}}{93.14~\mathrm{eV}} \: ,
\end{equation}
where we have introduced the effective mass $m_{s,\,\mathrm{eff}} \equiv m_s \Delta \Neff$. As mentioned in section~\ref{sec:intro}, 
light sterile neutrinos might behave as a hot dark matter component and affect the evolution of matter perturbations in a similar way as active neutrinos~\cite{Abazajian:2017tcc}. The form of the distribution function (\ref{eq:f_s}) implies that the sterile neutrinos have the same average momentum as the active species. Consequently, the maximum free-streaming length of the sterile is equal to the one of an active neutrino with mass $m_\nu=
m_s$, corresponding to a comoving wavenumber $k_\mathrm{fs}$ \cite{2013neco.book.....L}:
\begin{equation}
k_\mathrm{fs} = 0.018\,\Omega_m^{1/2} \left(\frac{m}{1\,\mathrm{eV}}\right)^{1/2} h \,\mathrm{Mpc}^{-1},
\end{equation}
if neutrinos become nonrelativistic during the matter-dominated era, or
\begin{equation}
k_\mathrm{fs} = 0.776\,\Omega_r^{1/2} \left(\frac{m}{1\,\mathrm{eV}}\right)^{1/2} h \,\mathrm{Mpc}^{-1},
\end{equation}
if they become non-relativistic during the radiation-dominated era. Here $m$ can be either the mass of an active neutrino, or the mass of sterile neutrino produced through non-resonant oscillations. Their large velocities prevent neutrinos from clustering at scales smaller than the free-streaming length, so the collective effect of active and sterile neutrinos is a step-like suppression of the amplitude of matter perturbations below the free-streaming scale~\cite{Colombi:1995ze,2013neco.book.....L}. 
\begin{equation}
     \frac{P_\nu}{P} \stackrel{k \gg k_\mathrm{fs}}{\approx} 1 - 8 f_\mathrm{hot} \: ,
\end{equation}
where $P_\nu$ and $P$ refer to the matter power spectrum with or without neutrinos respectively. The size of the suppression at small scales depends on the fraction of matter density provided by neutrinos, $f_\mathrm{hot}=(\Omega_\nu + \Omega_s)/\Omega_m$. For sterile neutrinos the onset of the suppression is determined by the free-streaming scale and therefore $m_s$, while the size of the suppression is proportional to $m_{s,\,\mathrm{eff}} = m_s \Delta N_\mathrm{eff}$. The cosmological effects of the sterile neutrino are therefore completely determined
once $m_s$ (or equivalently $\Omega_s$) and $\Delta \Neff$ are specified.

\begin{figure}
\centering
\includegraphics[width=1\textwidth]{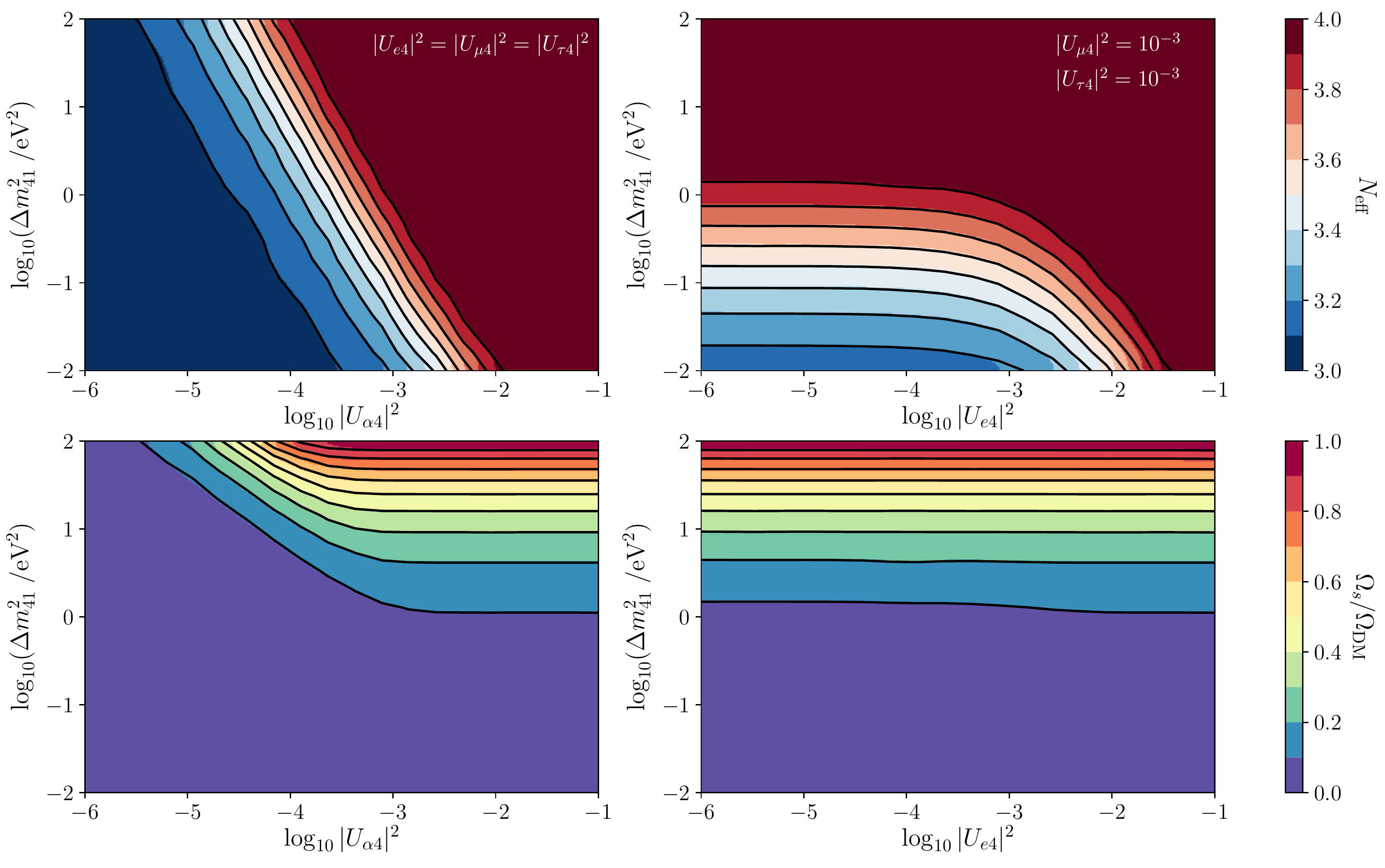}
\caption{\label{fig:theoretical_Neff_Omegas}
\textbf{Top:} $\Neff$ as a function of the mixing parameters $\left(\dmij{41}, \Uaj{e4}, \Uaj{\mu4}, \Uaj{\tau4} \right)$,
considering three equal varying mixing matrix elements ($\Uaj{e4}=\Uaj{\mu4}=\Uaj{\tau4}$) in the left panel
or one varying ($\Uaj{e4}$) and two fixed ($\Uaj{\mu4}=\Uaj{\tau4}=10^{-3}$) matrix elements in the right panel.
\textbf{Bottom:} Fraction of sterile dark matter compared to the total dark matter density
$\Omega_\mathrm{DM} h^2 = 0.119$ for the same mixing matrix elements assumptions.
}
\end{figure}

In figure~\ref{fig:theoretical_Neff_Omegas} we show the predicted contribution of the sterile neutrino
to \Neff\ (upper row)
and
the fraction of total dark matter energy density it represents
(lower row, assuming a total $\Omega_ch^2 = 0.119$ from the latest Planck measurements~\cite{Aghanim:2018eyx}), for two different choices for the mixing matrix elements.
From the top row of Fig.~\ref{fig:theoretical_Neff_Omegas}, it is clear that if at least one of the mixing matrix elements is much larger than $10^{-3}$, one has $\Delta\Neff~\simeq1$ for $m_s\sim1\,\mathrm{eV}$ in agreement with the expectation from Eq.~(\ref{eq:deltaneff_mixing_approx}) and the results
of previous analyses of active-sterile neutrino oscillations in the early Universe (see e.g.\ \cite{Hannestad:2012ky,Gariazzo:2019gyi}).
Such a value of $\Delta\Neff$ is at odds with what is inferred from Planck measurements of CMB anisotropies, or from astrophysical determinations of the abundances of light elements. In general, we note that in some parts of the parameter space probed by our analysis, corresponding to large mixing angles and/or large masses, the contribution of sterile neutrinos to both $\Neff$ and the dark matter energy density is large. The latter is also a problem, since cosmological observations also constrain the fraction of hot dark matter to be small. As we shall see in the next section, these regions of parameter space will be indeed excluded by data, in agreement with our expectations.


As we will discuss in detail in section~\ref{sec:results},
large contributions to $\Neff$ are in tension with cosmological constraints, so in
order to allow a sterile neutrino to form all the dark matter
one needs to move to higher masses $m_s \sim \mathrm{keV}$ and very small mixing angles. 
In this mass range, a sterile neutrino produced by oscillations would behave as warm (or even cold) dark matter, so limits on the hot component would not apply. Then even modest contributions to $\Neff$
are sufficient to provide the required dark matter density, although other astrophysical limits apply, as reviewed for instance in~\cite{Adhikari:2016bei,Abazajian:2017tcc,Boyarsky:2018tvu}.
The analysis of the mass range $\dmij{41}>10^2$ eV$^2$, however, is beyond the scope of this paper, as we are interested in light sterile neutrinos motivated by anomalies in neutrino oscillation experiments. The main question we want to address here is whether such a sterile state is compatible with cosmological bounds.


\section{Datasets and methods}
\label{sec:data}

In this section, we present the data sets employed in this analysis, and we discuss the method adopted in this work, including the set up adopted for the \texttt{FortEPianO} code. Since we perform our analysis in a Bayesian framework, we also report the prior choices for the parameters varied in the analysis. 

\subsection{Cosmological data}
\label{subsec:cosmodata}

We consider measurements of the CMB anisotropies in temperature and polarization~\cite{Aghanim:2019ame},
as well as determinations of the CMB lensing power spectrum \cite{Aghanim:2018oex},
as provided by the latest Planck 2018 data release~\cite{Akrami:2018vks,Aghanim:2018eyx}. The data sets are incorporated by using the publicly available \texttt{clik} likelihood code described in \cite{Aghanim:2019ame}.

To the CMB dataset, we add late-time distance and expansion rate baryon acoustic oscillation (BAO) measurements
from the 6dFGS~\cite{Beutler:2011hx}, SDSS-MGS~\cite{Ross:2014qpa}, and BOSS DR12~\cite{Alam:2016hwk} surveys. These measurements are fully consistent with the Planck results and greatly help to break parameter degeneracies.

\subsection{Oscillation experiments}
\label{subsec:oscidata}

As described in section~\ref{subsec:global_fits}, various oscillation experiments find conflicting results for the active-sterile mixing parameters and it is not possible to combine the measurements in a self-consistent way. Therefore we group the non-conflicting oscillation measurements and present their results separately.

The $\bar \nu_e$ flux from nuclear reactors is measured by Bugey-3 \cite{Achkar:1996rd},
DANSS \cite{Alekseev:2018efk,danilov_epshep19}
NEOS \cite{Ko:2016owz},
and
PROSPECT \cite{Ashenfelter:2018iov}. Note that these experiments use differential measurements at a varied distance from the source, so a calibration of the intrinsic neutrino flux is not necessary. A combined frequentist analysis results in a preference for non-zero mixing with a sterile at about $\sim 2.5 \sigma$ with two nearly degenerate best-fit points, located at
($\dmij{41}\simeq 0.4~$eV$^2$, $\Uaj{e4}\simeq0.01$)
and
($\dmij{41}\simeq1.3~\mbox{eV}^2$, $\Uaj{e4}\simeq0.01$)
\cite{sterile19} (see also \cite{Gariazzo:2018mwd,Dentler:2017tkw,Dentler:2018sju} for previous analyses).


We also include bounds on the mixing with muon neutrinos from measurements of the atmospheric flux by IceCube \cite{Aartsen:2017bap,TheIceCube:2016oqi} and from the MINOS+ experiment \cite{Adamson:2017uda}, which uses an accelerator beam with two detectors, one close ($\sim 500$~m) and one far ($\sim 800$~km) from the source, to put strong constraints covering a wide range of $\dmij{41}$ and $\Uaj{\mu 4}$. Both are combined in a frequentist analysis \cite{sterile19} and we refer to the combination of both data sets as ``$\nu_\mu$ disappearance''. Neither IceCube nor MINOS+ find anomalous events and therefore provide upper bounds on the mixing parameters. IceCube also has a limited sensitivity on $\Uaj{\tau 4}$ thanks to the low-energy data from DeepCore \cite{Aartsen:2017bap}.

\subsection{Decay experiments}
\label{subsec:decaydata}

In this work, we consider the latest measurements of the tritium $\beta$ decay spectrum released by the KATRIN collaboration~\cite{Aker:2019uuj} that sets a limit $m_\beta<1.1\,\mathrm{eV}$ at 90\% C.L. 
We include the constraint by using the approximated analytical likelihood function presented in Eq.~(B.3) of Ref.~\cite{Huang:2019tdh}. Since the modification to the decay energy spectrum induced by an extra light sterile species is below the resolution of KATRIN, this likelihood is also valid for the additional sterile neutrino.

Concerning $\nbb$ searches, we consider the results from the KamLAND-ZEN collaboration phase-I~\cite{PhysRevLett.110.062502},
and phase-II~\cite{PhysRevLett.117.082503},
from the GERDA collaboration~\cite{Agostini:2017iyd},
and from the EXO collaboration~\cite{EXO200}.
The constraints are included by using the approximated likelihoods given in \cite{Caldwell:2017mqu}.
A summary of all data sets used, the isotopes employed by each collaboration and the respective bounds on the lifetime $T_{1/2}$ can be found in table~\ref{tab:0nubb_experiments}, and we show the limits on $m_{\beta \beta}$ from the individual experiments in appendix~\ref{sec_app:individual_0nubb}.

While this paper was in preparation, new results from $0\nu\beta\beta$ collaborations have been published from the GERDA collaboration~\cite{Agostini:2019hzm}, the EXO200 collaboration~\cite{PhysRevLett.123.161802}, the CUORE collaboration~\cite{Adams:2018yrj}. We note that the tightest constraints on $m_{\beta \beta}$ are still those by KamLAND-ZEN phase-II. As a result, the inclusion of other released data from $0\nu\beta\beta$ searches would not change the conclusions of this work.


\begin{table}[tbp]
\centering
\renewcommand{\arraystretch}{1.3}
\begin{tabular}{|l c c c c|}
\hline
Collaboration &  Isotope &$T_{1/2}\,[10^{25}\,\mathrm{yr}]$ & $m_{\beta \beta} \, [\mathrm{eV}]$ & Ref.\\
\hline
GERDA &  $^{76}\mathrm{Ge}$ & $>5.3$ & $< 0.25$& \cite{Agostini:2017iyd}\\ 
KamLAND ZEN I & $^{136}\mathrm{Xe}$ & $>1.9$ & $< 0.21$ &\cite{PhysRevLett.110.062502}\\
KamLAND ZEN II & $^{136}\mathrm{Xe}$ & $>10.7$ & $< 0.10$ &\cite{PhysRevLett.117.082503}\\ 
EXO200 & $^{136}\mathrm{Xe}$ & $>1.1$ & $<0.24$  &~\cite{EXO200}\\ 
\hline
\end{tabular}
\caption{\label{tab:0nubb_experiments} Bounds on the half-life for $0 \nu \beta \beta$ events as measured by the various experiments we consider in our analysis, following~\cite{Gariazzo:2018pei,Caldwell:2017mqu}. The upper limits on $m_{\beta \beta}$ depend on the uncertainty of the nuclear matrix element and are not used directly in our analysis. Details on the individual limits and the effect of the nuclear matrix elements are summarized in appendix~\ref{sec_app:individual_0nubb}.
}
\end{table}

\subsection{Method}
\label{subsec:method}

In order to use cosmological data to constrain the sterile neutrino properties, we have to map the fundamental parameters $\left(\Delta m_{41}^2, |U_{e4}|^2, |U_{\mu4}|^2, |U_{\tau 4}|^2 \right)$ onto $\Neff$ and $m_s$, which determine its impact on cosmological observables.
As described in section~\ref{sec:sterile_cosmology} we use \texttt{FortEPiaNO} to calculate the evolution of the sterile distribution function including mixing with all three active neutrinos.
We pre-compute a $\Neff$ table on a four-dimensional grid spanning the parameter space we are interested in.
For the mass splitting, we consider nine logarithmically spaced samples over the range $\Delta m_{41}^2 \in [10^{-2}, 10^2]$~eV$^2$,
while for each mixing matrix element we take eleven logarithmically spaced samples in the range $\Uaj{\alpha 4} \in [10^{-6}, 10^{-1}]$.
The lower boundary chosen for the mass splitting ensures that the hierarchy $\Delta m_{21}^2 \ll \Delta m_{31}^2 \ll \Delta m_{41}^2$ always holds, so the sterile does not affect the oscillation patterns among active neutrinos which we keep fixed to their standard values. As a base model we assume a $\Lambda$CDM cosmology,
with its six parameters\footnote{The $\Lambda$CDM six parameters are the energy density in baryonic matter $\Omega_b h^2$ and in cold dark matter $\Omega_c h^2$, the angular size of the sound horizon at recombination $\theta$, the reionization optical depth $\tau$, the amplitude $A_s$ and spectral index $n_s$ of the primordial spectrum of scalar perturbations.},
complemented with the lightest neutrino mass $m_1$, the sterile mass splitting $\Delta m_{41}^2$
and the three mixing matrix elements \Uaj{\alpha4}, $\alpha=e,\,\mu,\,\tau$.

We calculate the evolution of cosmological perturbations
using the numerical Einstein-Boltzmann solver
\texttt{CLASS}~\cite{Lesgourgues:2011re,Blas:2011rf}~\footnote{\url{https://class-code.net/}}. Since we map the four active-sterile mixing parameters into the two cosmological parameters $\Neff$ and $m_s$,
a very large number of samples are necessary to obtain
a well-converged posterior in the full sterile parameter space.
To speed the calculations up, we therefore proceed in two steps.
First we use \texttt{MontePython}~\cite{Audren:2012wb,Brinckmann:2018cvx} to run a standard Markov Chain Monte Carlo (MCMC) with CMB+BAO data. For this run, we vary all $\Lambda$CDM parameters and include three massless neutrinos with $\Neff = 3.045$ and one additional massive state with varying mass and $\Delta \Neff$.
Note that with the current sensitivities the data cannot distinguish between various combinations of neutrino parameters as long as they result in the same $\Delta \Neff$ and $\mnu$,
so distributing the mass in different ways over the four available states does not affect the resulting constraints in a significant way (this reflects the fact that cosmology is not yet able to constrain the mass hierarchy, see e.g.~\cite{Hannestad:2016fog,Giusarma:2016phn,Gerbino:2016sgw,Gerbino:2016ehw,DiValentino:2016foa,Vagnozzi:2017ovm, Mahony:2019fyb}).
Given the prior $\Neff > 3.045$, the cosmological data sets a $95\%$-limit of $\mnu < 0.15$ eV and $\Neff < 3.44$.
We then use a Gaussian kernel density estimate (KDE) to interpolate the marginalized $\Neff - \mnu$ likelihood surface. Evaluating this estimated likelihood is $\sim 10^5$ faster than running the full CMB likelihood, and we explicitly make sure that sampling from it results in the same constraints as obtained from the MCMC.

In a second step, we use \texttt{emcee}~\cite{ForemanMackey:2012ig} together with the estimated likelihood to sample the neutrino parameter space as follows:
we vary the lightest neutrino mass $m_1$, the mass splitting $\dmij{41}$ and the three mixing matrix elements $\Uaj{\alpha 4}$. From the last four, we interpolate $\Delta \Neff$ from the pre-computed \texttt{FortEPiaNO} grid. Since the mass splittings $\dmij{21}$ and $\dmij{31}$ are much smaller than the sensitivity of current constraints on neutrino masses, we assume three degenerate active neutrinos $m_1 = m_2 = m_3$ so we get a total $\mnu = 3 m_1 + \Delta \Neff \, m_4$. We then evaluate the KDE of the cosmological likelihood for $\Neff = 3.045 + \Delta \Neff$ and $\mnu$. In order to compare the constraints to direct experiments, we also vary the nuclear matrix elements for $^{136}$Xe and $^{76}$Ge defined in Eq.~(\ref{eq:halflife}) and the three Majorana phases $\alpha_i$ in Eq.~(\ref{eq:m_bb}) to derive limits on $m_\beta$ and $m_{\beta \beta}$. A summary over all non-$\Lambda$CDM parameters varied in the analysis, their prior bounds and shapes is given in table~\ref{tab:priors}.

\begin{table}[tbp]
\centering
\renewcommand{\arraystretch}{1.3}
\begin{tabular}{|c c c c |}
\hline
\multicolumn{2}{|c}{Parameter} &  prior shape & prior bounds \\
\hline
lightest neutrino mass  & $m_1/\mathrm{eV}$ & flat & $[0, 5]$ \\
mass splitting & $\Delta m_{41}^2/\mathrm{eV}^2$ & log & $[10^{-2}, 10^2]$ \\
mixing matrix elements & $|U_{\alpha 4}|^2$ ($\alpha \in [e, \mu, \tau]$) & log & $[10^{-6}, 10^{-1}]$ \\
Majorana phases & $\alpha_j$ ($j \in [1, 2, 3]$)& flat & $[0, 2 \pi]$ \\
nuclear matrix elements & $|M_{0 \nu} ( ^{136}\mathrm{Xe}) |^2$ & flat & $[2.74, 3.45]$ \\
 & $|M_{0 \nu} ( ^{76}\mathrm{Ge}) |^2$  & flat & $[4.07, 4.87]$ \\
\hline
\end{tabular}
\caption{\label{tab:priors} Priors for all non-cosmological parameters varied in our analysis. We comment on different prior choices for $\Delta m_{41}^2$ in detail in section~\ref{subsec:results_priors}.}
\end{table}

For constraints involving the $\beta$-decay and $0\nu\beta\beta$ likelihoods described in section~\ref{subsec:decaydata}, we marginalize over all parameters listed in table~\ref{tab:priors} as well. The constraints from reactors and $\nu_\mu$ appearance measurements from MINOS+ and IceCube (see section~\ref{subsec:oscidata}), on the other hand, are derived in a frequentist framework and do not depend on the prior choices made here.


\section{Results}
\label{sec:results}

In this section we present the results of our analysis. First we discuss the complementarity of the various data sets either in terms of the fundamental parameters of the sterile mixing matrix, or in terms of the derived parameters $m_\beta$ and $m_{\beta \beta}$ directly constrained by terrestrial experiments. Then we explore the effect of assuming other priors on the sterile mass parameter for the cosmological datasets. In the same context we compare our constraints with the ones previously obtained in a simplified scenario where the sterile is only coupled to one neutrino species.

\subsection{Constraints from cosmology and direct experiments}
\label{subsec:results_compared}

\begin{figure}
\centering 
\includegraphics[width=1.\textwidth]{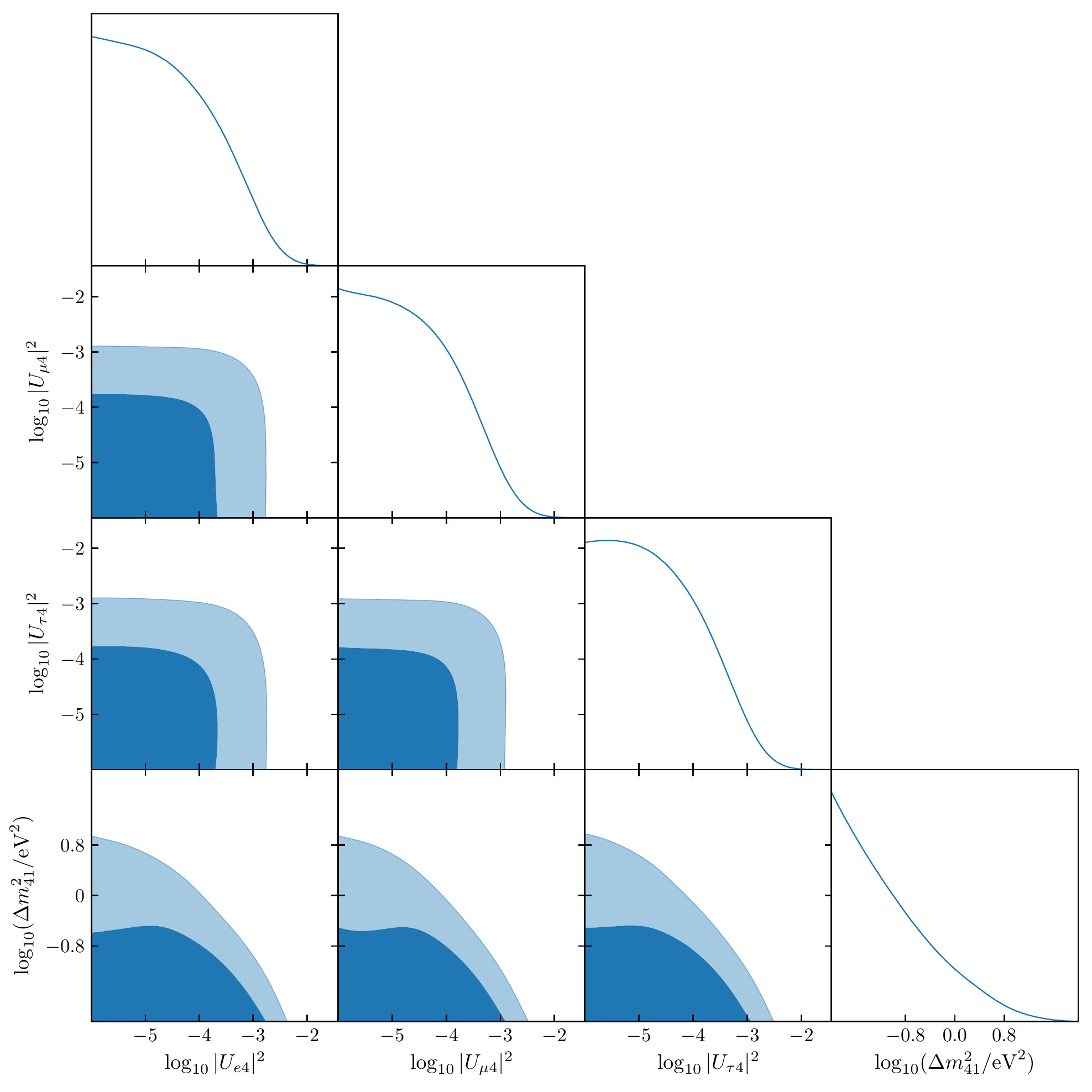}
\caption{\label{fig:cosmology_results}
Cosmological marginalized constraints on the mixing matrix elements $\Uaj{\alpha4}$ and mass splitting $\dmij{41}$ from CMB+BAO. Off-diagonal panels show  $68 \%$ and $95 \%$ confidence level probability contours. Panels along the diagonal show one-dimensional probability distributions. Cosmological constraints are not very sensitive to the flavor, so bounds on all different matrix elements are similar.
}
\end{figure}

Cosmology provides tight limits on the sum of neutrino masses $\mnu$ and the effective number of relativistic species $\Neff$,
that can be translated to tight bounds on the parameter space of the mixing matrix with significant relativistic energy contributions from the sterile neutrino.
After marginalising over all other cosmological and nuisance parameters, we obtain the constraints on the mixing matrix elements in figure~\ref{fig:cosmology_results} from CMB+BAO. As any mixing reaching a value of $|U_{\alpha4}|^2 \approx 10^{-3}$ starts to populate the sterile state and leads to detectable $\Neff$ contributions as seen in figure~\ref{fig:theoretical_Neff_Omegas}, this is where cosmological data becomes constraining.
There are only small differences between mixing for the various flavors, so the limits on all of the matrix elements are very similar.
Since the constraints on both the mass $m_4$ derived from the total sum $\mnu = 3 m_1 + \Delta \Neff \, m_4$ and $\Neff$ vanish for small amplitudes of the sterile distribution function set by $\Delta \Neff$, the cosmological data in principle allow large sterile masses as long as the mixing is small and the sterile state is not thermalized or populated to a significant level.
As mentioned in section~\ref{sec:sterile_cosmology}, much higher mass ranges in the keV range and larger are in principle possible if the mixing angles are small enough. In such a case the free-streaming length $k_\mathrm{fs}$ of the sterile neutrino would be pushed to smaller scales, and it would form a warm or even a cold dark matter component~\cite{Adhikari:2016bei,Abazajian:2017tcc,Boyarsky:2018tvu}.

\begin{table}[tbp]
\centering
\renewcommand{\arraystretch}{1.3}
\begin{tabular}{|c c c c c|}
\hline
\multirow{2}{*}{Parameter} & \multirow{2}{*}{experimental upper limit ($95\%$)} & \multicolumn{3}{c|}{cosmological upper limit ($95\%$)}  \\
 & & $\mathcal{P}(\log \dmij{14})$  &  $\mathcal{P}(m_4)$ &  $\mathcal{P}(\dmij{14})$ \\
\hline
$m_4$ [eV] & - & $1.6$ & $4.4$ & $6.8$\\
$\log_{10} \Uaj{e4}$& - & $-3.04$& $-3.43$& $-4.0$\\
$\log_{10}\Uaj{\mu4}$& $-2.2$ ($\nu_\mu$) & $-3.17$ &$-3.55$ & $-4.16$\\
$\log_{10}\Uaj{\tau 4}$& $-0.8$ ($\nu_\mu$)& $-3.18$& $-3.55$ &$-4.19$\\
$m_\beta$ [eV] & $0.9$ (KATRIN)&\multicolumn{3}{c|}{$0.09$}\\
$m_{\beta \beta}$ [eV] & $0.08$ ($0\nu \beta \beta$) &\multicolumn{3}{c|}{$0.07$}\\
\hline
\end{tabular}
\caption{\label{tab:results} Upper limits ($95 \%$) for the sterile neutrino mass, the parameters of the mixing matrix, and $m_\beta$ and $m_{\beta \beta}$. For the limits from direct experiments we take a conservative approach and only consider probes that are not in tension with cosmology. For those, we quote the strongest bound on each parameter. We present cosmological limits for the different prior choices for the mass splitting either on $\log \dmij{41}$ used in section~\ref{subsec:results_compared}, or on either $m_4$ or $\dmij{41}$ discussed in section~\ref{subsec:results_priors}, but note that the prior choice barely affects the resulting constraints on $m_\beta$ and $m_{\beta \beta}$. 
}
\end{table}

\begin{figure}[tbp]
\centering 
\includegraphics[width=.48\textwidth]{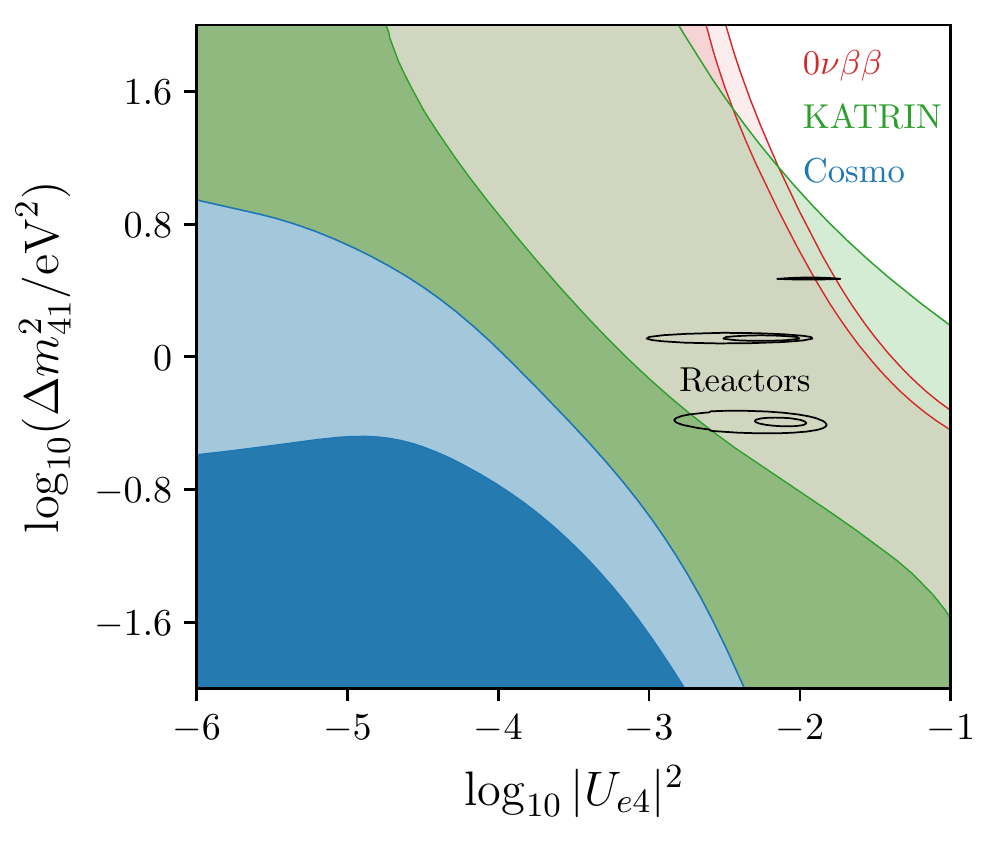}
\hfill
\includegraphics[width=.48\textwidth]{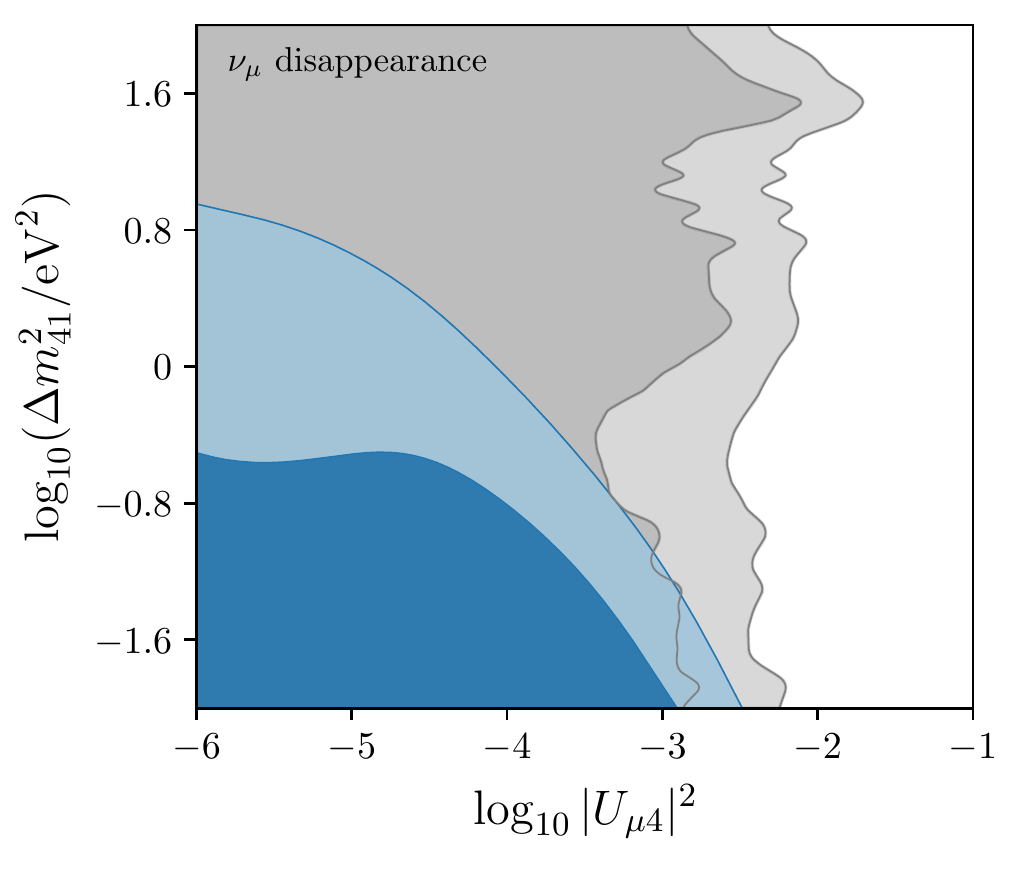}
\caption{\label{fig:0nuBB_CMB}
\textbf{Left:} Marginalized $68 \%$ and $95 \%$ constraints on the mass splitting \dmij{41} and
mixing matrix element \Uaj{e4} from cosmology (blue),
from the tritium $\beta$-decay measurements by KATRIN (green)
and from neutrinoless double-$\beta$-decay experiments ($0\nu \beta \beta$, red),
compared with the preferred frequentist regions by the joint fit \cite{sterile19} of reactor experiments
discussed in section~\ref{subsec:oscidata}, which are in strong tension with cosmological bounds.
\textbf{Right:} Cosmological $68 \%$ and $95 \%$ marginalized constraints on the mixing matrix element \Uaj{\mu4}
compared to (frequentist) $\nu_\mu$ disappearance results \cite{sterile19}
from IceCube and MINOS+ (grey).
}
\end{figure}

In figure~\ref{fig:0nuBB_CMB} (left panel), we compare the cosmological results with limits obtained from neutrinoless double-$\beta$ decay and tritium decay measurements by KATRIN \cite{Aker:2019uuj} in the $\Delta m_{41}^2 - |U_{e4}|^2$ plane, since the decay experiments are not sensitive to the other matrix elements.
While sensitivity of the neutrinoless double-$\beta$ decay experiments and $\beta$-decay measurements from KATRIN on the sterile parameters are comparable, cosmological bounds are orders of magnitude stronger.
To address the sterile neutrino interpretation of the reactor anomaly, we also show the parameter space preferred by a combined fit of short-baseline measurements of the reactor antineutrino flux.
As explained in section~\ref{sec:intro},
such experiments observe a preference in favor of a sterile with $\dmij{41} \sim \mathcal O (\mathrm{eV}^2)$ and $\Uaj{e4} \approx 10^{-2}$.
While these parameter values are compatible with current measurements from $0\nu\beta\beta$ and KATRIN, they are in severe tension with the CMB+BAO data. A mixing of the size needed to explain the reactor data would be more than sufficient to bring the sterile in thermal equilibrium in the early universe. On the right of figure~\ref{fig:0nuBB_CMB} we show a similar comparison for the mixing angle $\Uaj{\mu 4}$ and present cosmological constraints together with bounds from $\nu_\mu$ disappearance measurements from IceCube and MINOS+ described in section~\ref{subsec:oscidata}. While the experimental sensitivity is stronger than on $\Uaj{e 4}$, we still find the CMB+BAO data set to be more constraining.
While the cosmological limits rely on model assumptions and can be slightly relaxed in extended parameter spaces, accommodating $\Neff \approx 4$ is very challenging \cite{DiValentino:2019dzu, Kreisch:2019yzn}.

\begin{figure}
\centering 
\includegraphics[width=.9\textwidth]{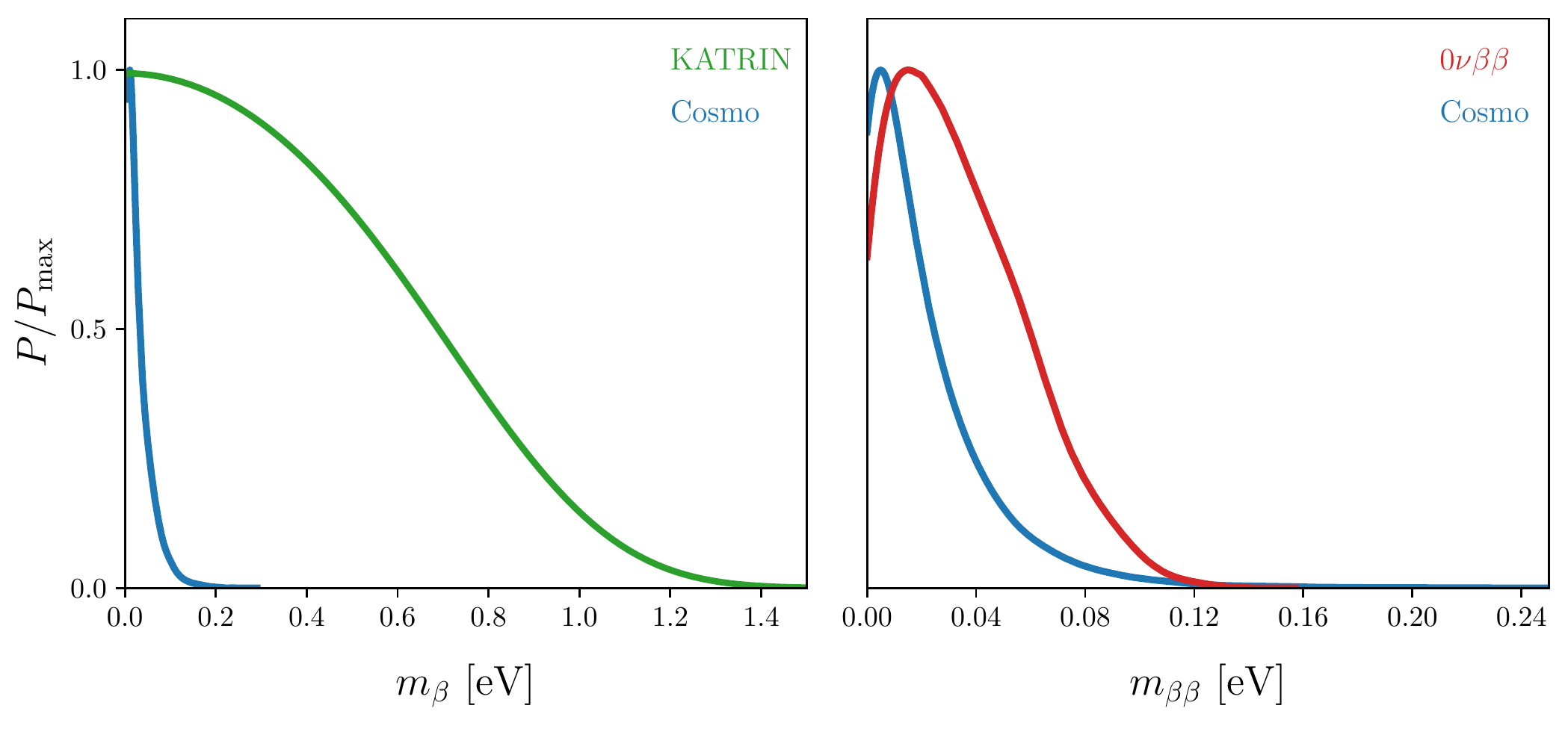}
\caption{\label{fig:mb_mbb} Left: One-dimensional probability distribution for the neutrino effective mass $m_\beta$ from cosmology and $\beta$-decay measurements from KATRIN. Right: One-dimensional probability distribution for the mass parameter $m_{\beta \beta}$ from $0 \nu \beta \beta$ and CMB+BAO. The combined $0\nu\beta\beta$ probability distribution is peaked at slightly non-zero values of $m_{\beta \beta}$ due to a modest excess of events observed at EXO200.}
\end{figure}

We also map the cosmological bounds onto the parameter space $m_\beta$ or $m_{\beta \beta}$ directly probed by decay experiments in figure~\ref{fig:mb_mbb}. On the left side we show limits from CMB+BAO together with the latest results from KATRIN~\cite{Aker:2019uuj}.
Since $m_\beta$ defined in Eq.~(\ref{eq:m_b}) receives contributions from the sterile, the resulting limit from cosmology in the extended $3+1$ parameter space is slightly higher compared to the expectation $m_\beta \approx m_1 \approx \mnu / 3$ from standard neutrinos.
The CMB+BAO data together with the prior assumption $\Neff > 3.045$ lead to an upper bound on the neutrino mass sum of $\mnu < 0.15$ eV, corresponding to a limit $m_\beta^\mathrm{cosmo} < 0.05$ eV assuming only three active neutrinos, which is slightly degraded to $m_\beta^\mathrm{cosmo} < 0.09$ eV if the sterile is included.
However, this constraint is still a factor of $\sim 10$ stronger than current experimental limits from KATRIN.

On the right-hand side of figure~\ref{fig:mb_mbb} we show a similar comparison of the derived bound on $m_{\beta \beta}$ from cosmology and direct $0\nu \beta \beta$ searches. Note that the posterior from the combined $0 \nu \beta \beta$ data has a maximum at slightly non-zero values due to an excess of events observed by EXO200 compared to the background expectation \cite{PhysRevLett.123.161802}.
In table~\ref{tab:results}, we present a summary of all $95 \%$ bounds on the sterile mass $m_4$, the mixing matrix elements $\Uaj{\alpha4}$ for each flavor, $m_\beta$ and $m_{\beta\beta}$, comparing the cosmological bound to the respective strongest bound from direct searches. Cosmology provides the tightest limits on all parameters, for most of them by orders of magnitude.



\subsection{Priors and parameter space volume effects}
\label{subsec:results_priors}

Since all cosmological bounds presented in figure~\ref{fig:cosmology_results} and discussed in the previous section~\ref{subsec:results_compared} provide only upper bounds on the sterile parameters in a Bayesian framework, the choice of priors affects the limit.
In this section we focus on various priors for the mass splitting $\Delta m_{41}^2$.
While we adopt a logarithmic prior as a standard case, since cosmological data is ignorant of the order of magnitude of the mass splitting, one can also argue that the parameter of interest is either the mass-squared difference itself --- or the sterile mass scale, since cosmological data is approximately sensitive to the energy density parameter $\Omega_s \propto m_s \approx m_4$.
We therefore repeat the previous cosmological analysis using a flat prior either on $\log_{10} \dmij{41}$, on $\dmij{41}$, or on $m_4$, always considering the same range specified before in table~\ref{tab:priors}.
Note that while sampling over $m_4$ we enforce the additional constraint $\dmij{41} > 10^{-2} \, \mathrm{eV}^2$ to make sure that the neutrino mass hierarchy is unchanged.

\begin{figure}
\centering 
\includegraphics[width=1.\textwidth]{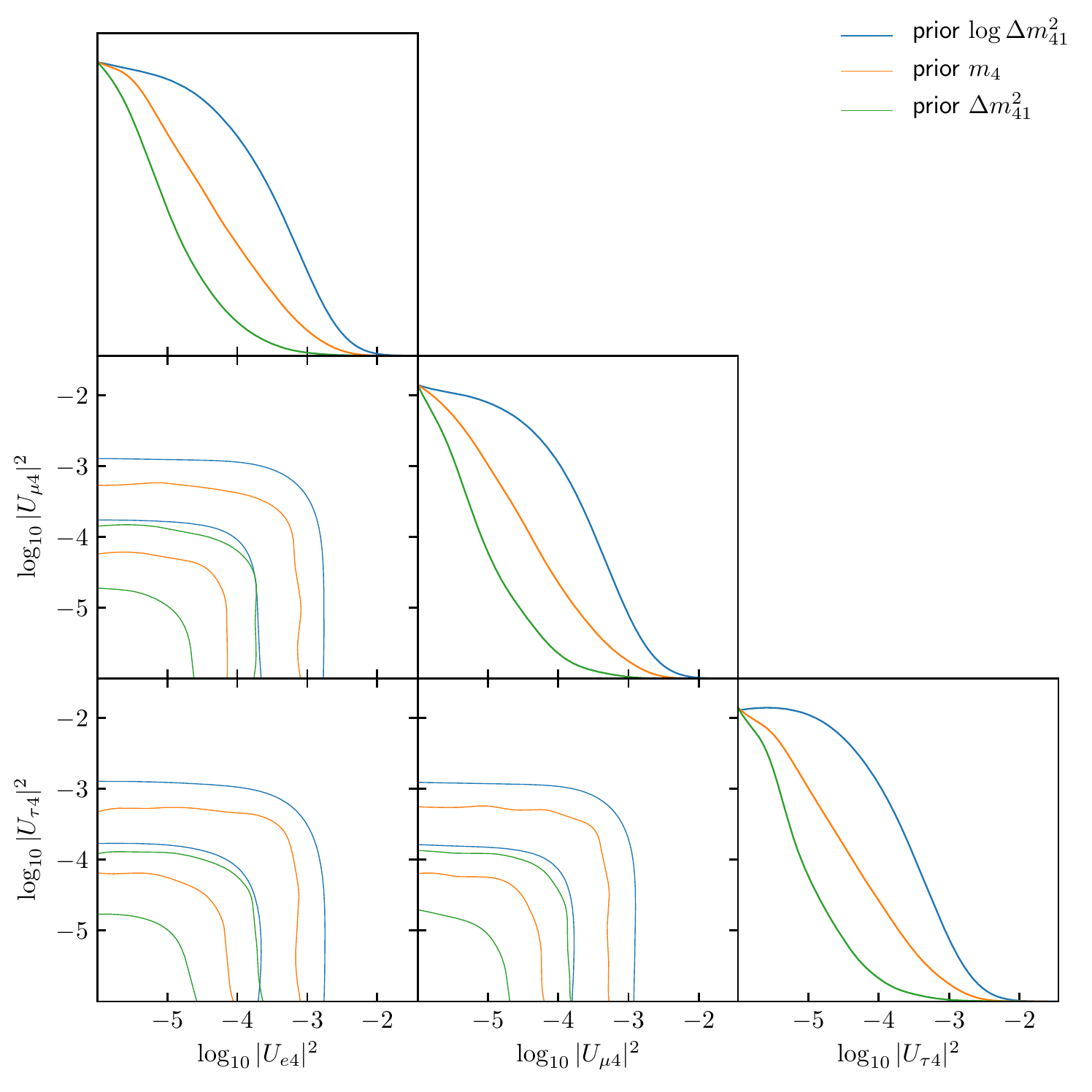}
\caption{\label{fig:priors}
Cosmological marginalized constraints on the mixing matrix elements $\Uaj{\alpha4}$ for flat priors on either $\log \Delta m_{41}^2$ (blue), $m_4$ (orange) or $\Delta m_{41}^2$ (green), from CMB+BAO data.
Off-diagonal panels show  $68 \%$ and $95 \%$ confidence level probability contours.
Panels along the diagonal show one-dimensional probability distributions.
The more weight the prior gives to higher sterile masses, the lower the allowed $|U_{i4}|^2$ values in order to stay within the $\Neff$ range allowed by the cosmological data.
}
\end{figure}

We present the resulting limits on the mixing matrix elements in figure~\ref{fig:priors} and in table~\ref{tab:results}.
The different choices indeed affects the results, and bounds on the mixing matrix elements are stronger for the flat priors on $m_4$ or $\dmij{41}$.
This is a consequence of the mapping between parameter spaces: the most important constraint from the data is on $\Neff$ (which also contributes to $\mnu$ through $\Delta \Neff \, m_4$), and there is a strong degeneracy between the mass splitting and the mixing matrix elements as explained in section~\ref{sec:ster_osc} and seen from the upper left plot in figure~\ref{fig:theoretical_Neff_Omegas}.
For a fixed higher sterile mass splitting, the mixing angles have to be smaller to stay within the allowed $\Neff$ region. Therefore the more weight the priors give to large masses $m_4$ or $\dmij{41}$, the smaller the matrix elements $\Uaj{\alpha4}$ have to be to fulfill the $\Delta \Neff$ constraint. We want to emphasize that the choice of priors does not affect the degree of tension with reactor and $\bar \nu_e$ appearance measurements.
A different choice of priors changes the weight of the mapping $(\dmij{41}, \Uaj{\alpha 4}) \rightarrow \Neff$, but the region in parameter space necessary to explain the anomalies leads to an almost thermal sterile with $\Delta \Neff \approx 1$. This is excluded by cosmological data independent of the assumptions made in the analysis.

\begin{figure}[tbp]
\centering 
\includegraphics[width=.48\textwidth]{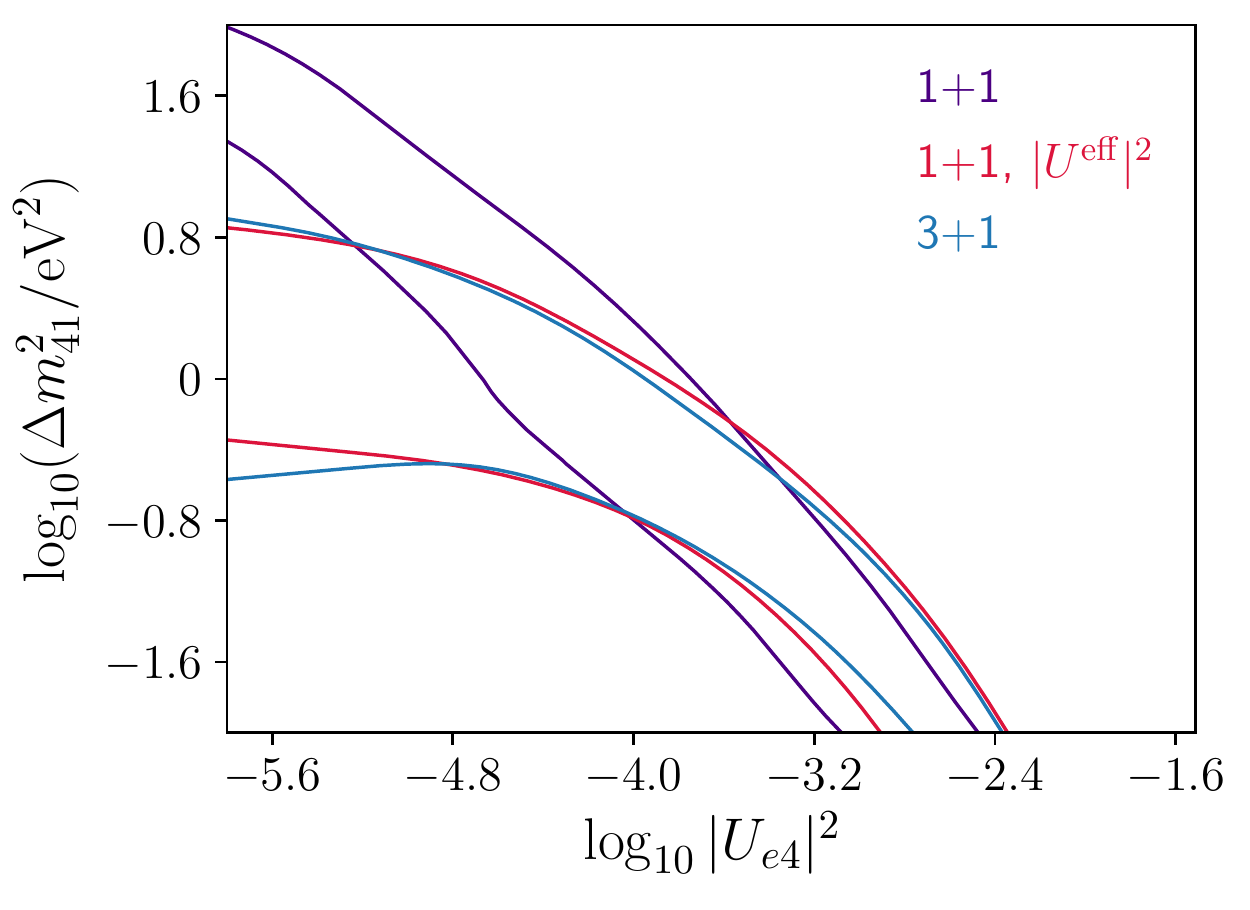}
\hfill
\includegraphics[width=.48\textwidth]{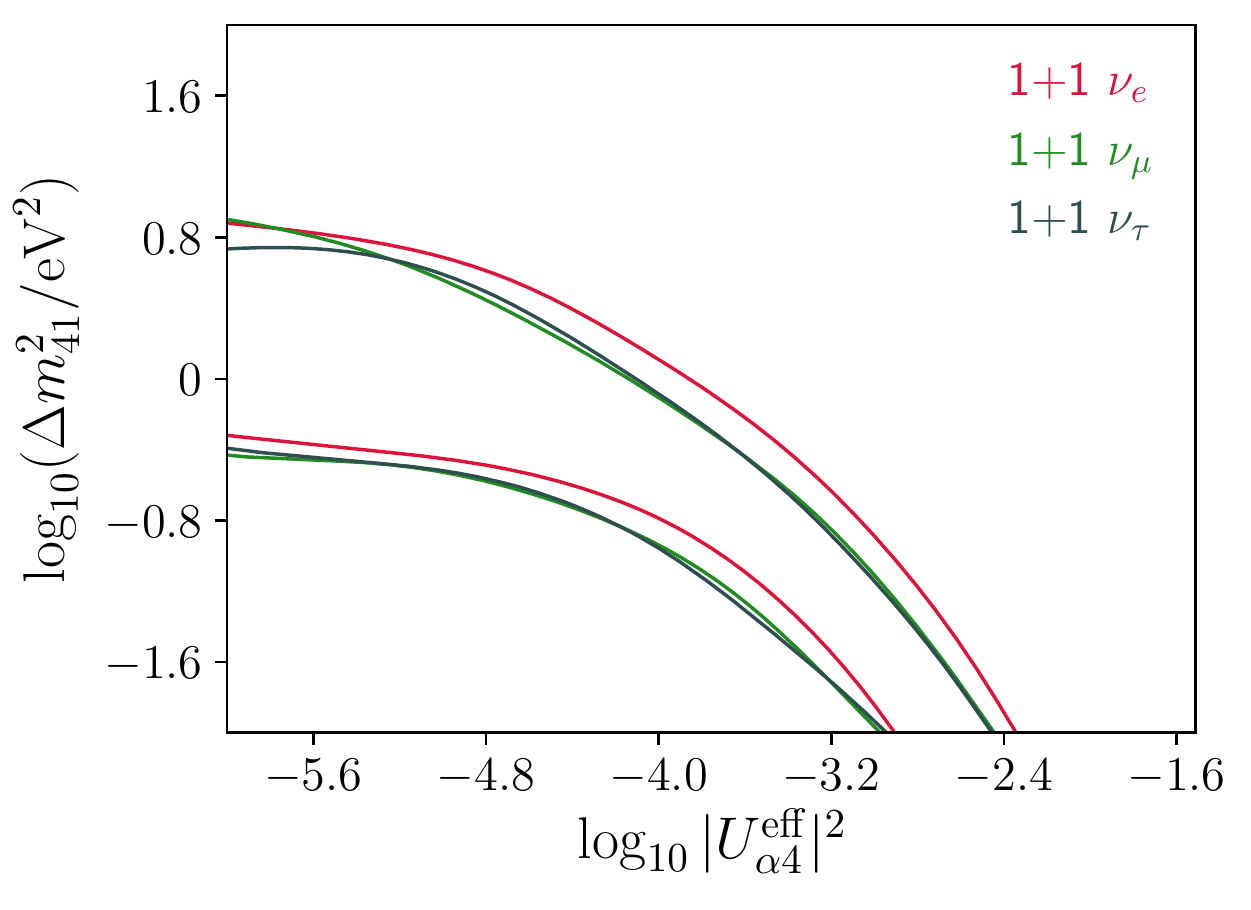}
\caption{\label{fig:11vs31} \textbf{Left:} Cosmological $68 \%$ and $95 \%$ marginalized constraints assuming either a $1+1$ case with the sterile only coupled to $\nu_e$ (purple), compared to the full $3+1$ scenario including the full mixing matrix (blue) from figure~\ref{fig:cosmology_results}. The difference is mostly a parameter space volume effect, since the limits obtained with a modified coupling only to $\nu_e$ with an effective mixing $|U^\mathrm{eff}|^2 = \Sigma_{\alpha} |U_{\alpha 4}|^2$ ($\alpha \in [e, \mu, \tau]$, red) are almost the same as for the $3+1$ case. \textbf{Right:} Cosmological $68 \%$ and $95 \%$ marginalized constraints on $|U^\mathrm{eff}|^2$ for a single sterile coupled only to one neutrino species $\nu_e$ (red, same as left), $\nu_\mu$ (green) or $\nu_\tau$ (grey). The differences between coupling to the different flavors are very small.}
\end{figure}

The resulting limits on the mixing matrix elements summarized in table~\ref{tab:results} are robustly constrained to be $\Uaj{\alpha4} < 10^{-3}$ for all flavors. While mixing with electron neutrinos is the most important channel and the resulting bounds on $\Uaj{e 4}$ are slightly more stringent, there is overall only a minor difference between mixing with the different active neutrinos.
It is therefore interesting to understand the difference between a simplified case where the sterile is only coupled to one active neutrino (often assumed to be $\nu_e$) and the full mixing with all flavors explored in this paper.
We present the comparison between the cosmological constraints on $\dmij{41}$ and $\Uaj{e4}$ assuming either a $1+1$ or a $3+1$ scenario on the left side of figure~\ref{fig:11vs31}.
The results clearly differ and the $1+1$ mixing allows for higher mass splittings of the sterile.
However, due to parameter space volume effects it is expected that the limits look different.
For every point in the $(\dmij{41}, \Uaj{e4})$ plane, in the $3+1$ scenario there are more ways to achieve a higher $\Delta \Neff$ and $\mnu$ by increasing the other mixing matrix elements.
Higher mass splittings with small $\Uaj{e4}$ that lead to negligible cosmological effects in the $1+1$ model are then still unlikely in the $3+1$ case since the other mixing angles have to be small as well.
We explicitly test this explanation by adding another comparison case where the sterile is still only coupled to electron neutrinos, but we consider an effective mixing matrix element
\begin{equation}
\label{eq:Ueff}
    |U^\mathrm{eff}|^2 = \sum_{i}^3 \Uaj{i4}
\end{equation}
and we sample over three distinct contributions $\Uaj{i4}$ to make up for the larger parameter space volume in the full $3+1$ mixing scenario.
As can be seen on the left side of figure~\ref{fig:11vs31}, this parameter space volume effect almost completely accounts for the difference between the scenarios. On the right-hand side of figure~\ref{fig:11vs31} we test this effective mixing scenario by accounting for the parameter space effect in the same way, but coupling the sterile to either one of electron-, muon- or tau-neutrinos respectively. Again, differences between the individual flavors are very small, and an effective coupling to $\nu_e$ provides a very good approximation to the full dynamics.

As a consequence we find that the production of light sterile neutrinos via mixing with to three active states can be modelled within a $1+1$ model with a single sterile mixing with $\nu_e$ as long as the effective mixing matrix element in Eq.~(\ref{eq:Ueff}) is used. Therefore, the abundance of a single sterile computed in the extended 3+1 model is, to an excellent approximation, very similar to that found in the effective 1+1 scenario, provided that the 1+1 squared mixing matrix element is the sum of the individual squared mixing matrix elements in the 3+1 case.



\section{Conclusion}
\label{sec:conclusion}

Several anomalies observed in short-baseline oscillation experiments hint at a new neutrino mass state at the eV scale. In this paper, we have provided a consistent framework to constrain the additional neutrino mass splitting and the active-sterile mixing angles with cosmological data. This also allows us to compare the bounds from cosmological data with results from oscillations, $\beta$-decay and $\nbb$ measurements in a common parameter space.

For the first time, we performed this analysis in a full $3+1$ scenario where the sterile state is 
mixed with all three active neutrinos.
In order to map the sterile neutrino mass-squared splitting and mixing matrix parameters for each flavor $\dmij{41}$ and $\Uaj{\alpha4}$ onto the cosmological observables, we have solved the evolution of the $4 \times 4$ neutrino density matrix in the early universe with the \texttt{FortEPiaNO}~\cite{Gariazzo:2019gyi} code and find that the resulting distribution function of the fourth neutrino state is well approximated by a Dodelson-Widrow form \cite{Dodelson:1993je}.

A combination of CMB and BAO data currently provides the strongest available bounds on the sterile mass splitting $\dmij{41}$ and the mixing matrix elements $\Uaj{\alpha4}$. While the limit on the mass scale depends on prior assumptions, we robustly find that once any mixing matrix element reaches a level of $\Uaj{\alpha4} \approx 10^{-3}$, the new state would give rise to a detectable relativistic energy contribution in the early universe not seen in cosmological data. The parameter space needed to explain short baseline anomalies with a sterile neutrino leads to a fully thermalized relativistic species with $\Neff \approx 4$ and is in strong tension with the CMB bounds.
We also derive limits on the effective electron neutrino mass $m_\beta$ and the Majorana mass parameter $m_{\beta\beta}$, measurable in $\beta$- and $\nbb$-decay experiments, from cosmological data, finding $m_\beta<0.09$ eV and $m_{\beta \beta}<0.07$ eV at 95\% C.L. These constraints are tighter than the ones obtained from the latest direct laboratory measurements. Our main results in terms of limits on the sterile mass scale $m_4$, the mixing matrix parameters for each flavor $\Uaj{\alpha4}$, $m_\beta$, and $m_{\beta\beta}$, are summarized in table~\ref{tab:results}.

We have explored the effect of prior choices on the cosmological bounds repeating the analysis with different priors on the mass splitting. Since a sizeable sterile contribution to $\Neff$ can be produced either by a large mass splittings or large mixing angles, the main effect of priors is to shift the weight between the two quantities. The maximal contribution to $\Neff$ allowed by the data is fixed, so the more weight the prior gives to high mass splittings $\dmij{41}$, the lower the mixing angles have to be to stay within the allowed region.

Almost all limits on sterile neutrinos from cosmology previously reported in the literature were derived in a simplified $1+1$ scenario where the sterile is only coupled to one active neutrino. Even though the mixing with the different active flavors is almost equivalent, we show that the results differ significantly from the ones obtained with a $3+1$ mixing scheme assumed for this work. This can be largely attributed to parameter space volume effects and can be accounted for by coupling the sterile with one active neutrino with effective mixing as given by Eq.~(\ref{eq:Ueff}). In any case, our results show that allowing for more than one active-sterile mixing does not relax the tension between the active-sterile neutrino parameters favored by the oscillation anomalies and present cosmological observations.

Since cosmological data currently dominates the constraints, we do not perform a joint analysis with laboratory experiments at this time. However, the framework provided here can be the basis for new global constraints on light sterile neutrino properties once new data from laboratory searches becomes available.

\acknowledgments

SH, PFdS, and KF acknowledge support from the Vetenskapsr\r{a}det (Swedish Research Council) through contract No.\ 638-2013-8993 and the Oskar Klein Centre for Cosmoparticle Physics. SG acknowledges support from the European Union's Horizon 2020 research and innovation program under the Marie Sk\l odowska-Curie individual Grant Agreement No.\ 796941. SG and SP acknowledge support from the Spanish grants FPA2017-85216-P (AEI/FEDER, UE), PROMETEO/2018/165 (Generalitat Valenciana) and the Red Consolider MultiDark FPA2017-90566-REDC. MG acknowledges support from Argonne National Laboratory (ANL). ANL's work was supported by the DOE under contract W7405-ENG-36. MG and ML acknowledge support from the ASI grant 2016-24-H.0 COSMOS ``Attivit\`{a} di studio per la comunit\`{a} scientifica di cosmologia'' and from INFN through the InDark and Gruppo IV fundings. SV acknowledges support from the Isaac Newton Trust and the Kavli Foundation through a Newton-Kavli Fellowship, and acknowledges a College Research Associateship at Homerton College, University of Cambridge. KF acknowledges support from the Jeff and Gail Kodosky Endowed Chair in
Physics, DoE grant DE- SC007859 and the LCTP at the
University of Michigan.
This work is partly based on observations obtained with Planck (http://www.esa.int/Planck), an ESA science mission with instruments and contributions directly funded by ESA Member States, NASA, and Canada.

\appendix
\section{Individual $\nbb$ likelihoods}
\label{sec_app:individual_0nubb}

\begin{figure}
\centering
\includegraphics[width=0.6\textwidth]{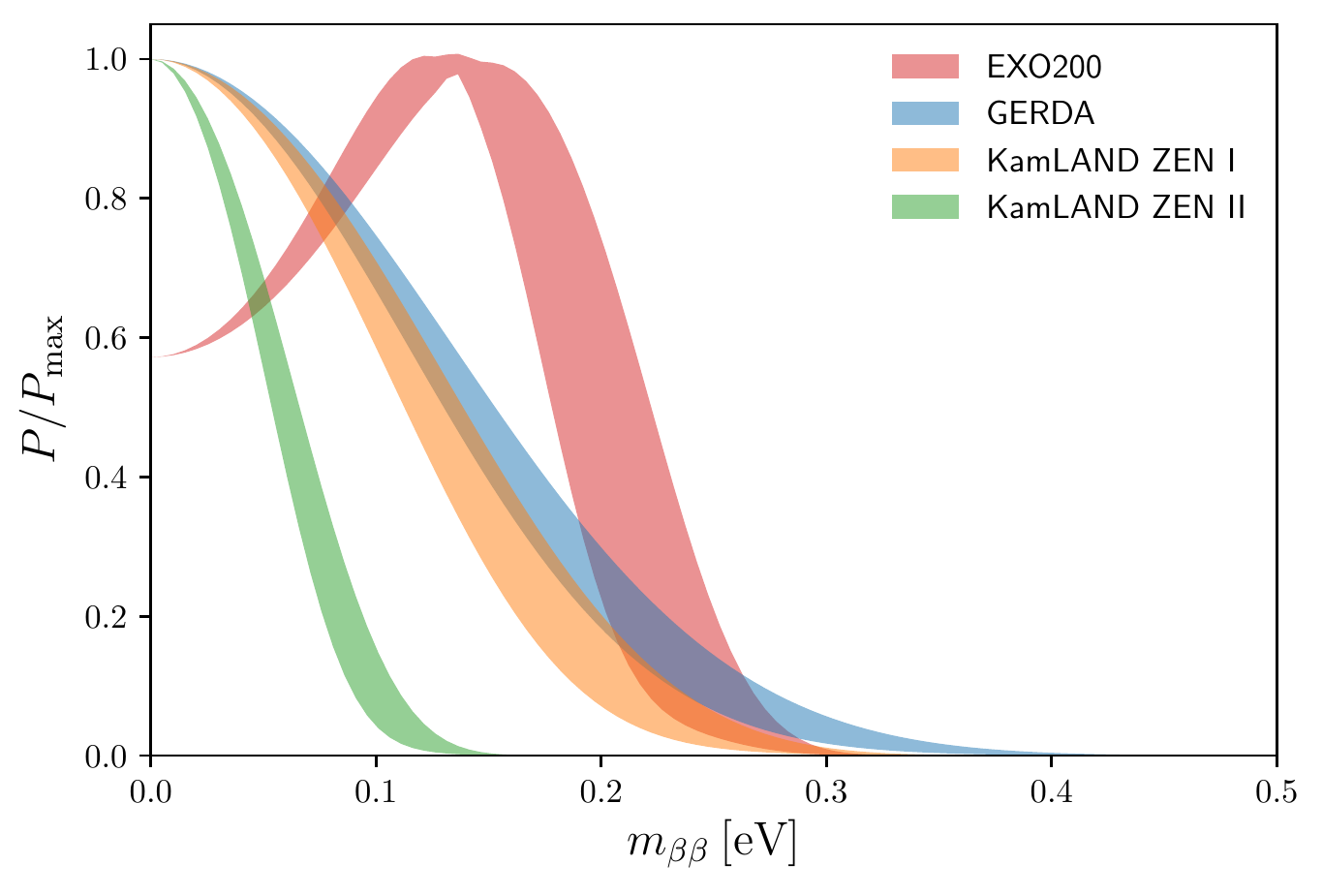}
\caption{\label{fig:individual_0nuBB}
Probability distributions for $m_{\beta \beta}$ from the individual $\nbb$ experiments used in this analysis based on the measured limit on $T_{1/2}$. The shaded region shows the result of varying the corresponding nuclear matrix transition element in Eq.~(\ref{eq:m_bb}).
}
\end{figure}

For completeness we also present the individual constraints on $m_{\beta \beta}$ from GERDA, KamLAND ZEN I \& II and EXO200 in figure~\ref{fig:individual_0nuBB}. The width of each curve represents the uncertainty in the nuclear matrix element given in table~\ref{tab:priors}. A modest excess of events observed at EXO200 makes the corresponding probability distribution for $m_{\beta\beta}$ to peak at slightly non-zero values of $m_{\beta\beta}$. The combination of the individual constraints presented in Fig.~\ref{fig:individual_0nuBB} gives the probability distribution reported as a red solid line labeled $0\nu2\beta$ in the right panel of Fig.~\ref{fig:mb_mbb}.

\bibliographystyle{JHEP}
\bibliography{Bibliography}

\end{document}